\documentclass[11pt,oneside]{article}
\pdfoutput=1
\usepackage{amsthm}
\usepackage{enumitem}
\usepackage{jheppub}
\usepackage[skip=2pt,font=footnotesize]{caption}
\usepackage[utf8]{inputenc}
\usepackage{verbatim}
\usepackage{amsmath}
\usepackage{amssymb}
\usepackage{amsthm}
\usepackage{slashed}
\usepackage{amsfonts}
\usepackage{makecell}
\usepackage{tabularx}
\usepackage{comment}
\usepackage{graphicx,subfigure}
\usepackage{float}
\usepackage{color}
\usepackage{psfrag}
\usepackage{color}
\definecolor{darkblue}{rgb}{0.1,0.1,.7}
\usepackage{hyperref} 

\makeatletter
\newcommand\xleftrightarrow[2][]{%
  \ext@arrow 9999{\longleftrightarrowfill@}{#1}{#2}}
\newcommand\longleftrightarrowfill@{%
  \arrowfill@\leftarrow\relbar\rightarrow}
\makeatother

\hypersetup{breaklinks}
\usepackage{comment}

\def\1{{\rm 1-loop}}

\usepackage{braket}

\def\<{\langle}
\def\>{\rangle}

\newcommand   \f  {\phi}

\newcommand{\bea}{\begin{eqnarray}}
\newcommand{\eea}{\end{eqnarray}}

\def\disc{\text{disc}}

\def\O{{\cal O}}

\newcommand   \llc {\<\!\<}
\newcommand   \rrc {\>\!\>}

\newcommand {\be} {\begin {equation}}
\newcommand {\ee} {\end {equation}}

\newcommand {\bes} {\begin {equation*}}
\newcommand {\ees} {\end {equation*}}

\DeclareMathOperator*{\res}{Res}

\newcommand{\beq}{\begin{equation}}
\newcommand{\eeq}{\end{equation}}

\def\be{ \begin{equation} }
\def\ee{ \end{equation} }

 \newmuskip\pFqmuskip

\usepackage{enumitem}

\newlist{primenumerate}{enumerate}{1}
\setlist[primenumerate,1]{label={3$'$.}}

\newcommand*\pFq[6][8]{%
  \begingroup 
  \pFqmuskip=#1mu\relax
  \mathcode`\,=\string"8000
  \begingroup\lccode`\~=`\,
  \lowercase{\endgroup\let~}\pFqcomma
  {}_{#2}F_{#3}{\left[\genfrac..{0pt}{}{#4}{#5};#6\right]}%
  \endgroup
}
\newcommand{\pFqcomma}{\mskip\pFqmuskip}

\makeatletter
\renewcommand{\@maketitle}{
\newpage
 \begin{center}%
  {\large\bfseries \@title \par}%
 \end{center}%
 \par} \makeatother

\numberwithin{equation}{section}

\begin{document}
\hfill \hbox{{CALT-TH-2021-028}}

\title{
The Inflationary Wavefunction from Analyticity and Factorization
}
\author{David Meltzer}

\affiliation{Walter Burke Institute for Theoretical Physics, California Institute of Technology,\\ 1200 East California Boulevard, Pasadena, California, 91125}

\emailAdd{dmeltzer@caltech.edu}

\abstract{ 
We study the analytic properties of tree-level wavefunction coefficients in quasi-de Sitter space.
We focus on theories which spontaneously break dS boost symmetries and can produce significant non-Gaussianities.
The corresponding inflationary correlators are (approximately) scale invariant, but are not invariant under the full conformal group.
We derive cutting rules and dispersion formulas for the late-time wavefunction coefficients by using factorization and analyticity properties of the dS bulk-to-bulk propagator.
This gives a unitarity method which is valid at tree-level for general $n$-point functions and for fields of arbitrary mass.
Using the cutting rules and dispersion formulas, we are able to compute $n$-point functions by gluing together lower-point functions.
As an application, we study general four-point, scalar exchange diagrams in the EFT of inflation.
We show that exchange diagrams constructed from boost-breaking interactions can be written as a finite sum over residues.
Finally, we explain how the dS identities used in this work are related by analytic continuation to analogous identities in Anti-de Sitter space.
}

\maketitle

\newpage
\section{Introduction}
\label{sec:intro}

According to inflation, the early universe went through a phase of quasi-de Sitter expansion that stretched quantum fluctuations to superhorizon scales \cite{Guth:1980zm,Linde:1981mu,Albrecht:1982wi,Starobinsky:1982ee}.
Current measurements show that inflationary correlators are nearly Gaussian and scale invariant \cite{Planck:2019kim}.
A discovery of non-Gaussianities in future experiments \cite{Meerburg:2019qqi} would be an exciting development as they encode information about the physics of inflation, its UV completion, and the possible existence of new particles \cite{Chen:2010xka,Arkani-Hamed:2015bza,Lee:2016vti}.
The goal of this work is to study inflationary correlators using a bootstrap approach that leverages their analytic properties.  

The natural observables in dS are equal-time correlation functions at future infinity.
In weakly-coupled theories one can compute these correlators by summing over all relevant Feynman diagrams.
However, even tree-level dS Feynman diagrams can be difficult to compute in practice.
Individual Feynman diagrams can also obscure symmetries that are manifest in the final answer.
Therefore, it is useful to understand how dS correlators can be studied from a purely boundary, or on-shell, perspective.
On-shell methods have yielded new insights into the mathematical structure of flat-space scattering amplitudes \cite{Elvang:2013cua,Cheung:2017pzi}, and it is natural to expect that similar progress can also be made for cosmological observables.
On-shell methods for inflation have recently been developed in the cosmological bootstrap program \cite{Arkani-Hamed:2017fdk,Arkani-Hamed:2018kmz,Baumann:2019oyu,Baumann:2020dch,Baumann:2021fxj,Sleight:2019mgd,Sleight:2019hfp,Sleight:2020obc,Sleight:2021iix,Green:2020ebl,Cespedes:2020xqq,Hillman:2019wgh,Goodhew:2020hob,Pajer:2020wxk,Jazayeri:2021fvk,Melville:2021lst,Goodhew:2021oqg,Bonifacio:2021azc}, which uses physical consistency conditions and symmetries to fix the late-time correlators.
In particular, these works have shown the importance of studying dS correlators as analytic functions of their momenta and imposing that they have the correct singularity structure.\footnote{For related work on simplifying AdS/CFT perturbation theory in momentum space see \cite{Raju:2010by,Raju:2011mp,Raju:2012zr,Raju:2012zs,Isono:2018rrb,Isono:2019wex,Albayrak:2018tam,Albayrak:2019asr,Albayrak:2019yve,Albayrak:2020isk,Albayrak:2020bso,Albayrak:2020fyp,Armstrong:2020woi,Costantino:2020vdu}.}

Late-time correlators in pure dS$_{d+1}$ are invariant under the $d$-dimensional, Euclidean conformal group, $\text{SO}(d+1,1)$. 
Conformal invariance imposes strong constraints on the form of boundary dS correlators \cite{Antoniadis:2011ib,Creminelli:2011mw,Maldacena:2011nz,Kehagias:2012pd,Mata:2012bx,Kundu:2014gxa,Kundu:2015xta,Ghosh:2014kba,Pajer:2016ieg}\footnote{One subtlety is that correlators may only be invariant under a combined conformal and coordinate transformation \cite{Ghosh:2014kba}.}, which can be used to bootstrap four-point functions from first principles \cite{Arkani-Hamed:2018kmz}.
There has also been progress bootstrapping theories which break dS boost symmetries, or boundary special conformal transformations \cite{Green:2020ebl,Goodhew:2020hob,Pajer:2020wxk,Jazayeri:2021fvk,Melville:2021lst,Goodhew:2021oqg,Baumann:2021fxj,Bonifacio:2021azc}.
This generalization is important because inflaton backgrounds generically break dS boost symmetries by an order one amount.
Moreover, only boost-breaking interactions can produce large non-Gaussianities in single-clock theories \cite{Green:2020ebl}.
Bulk time translations, or boundary scale transformations, are also broken by the inflaton background, but this symmetry can be (approximately) restored if the scalar sector of the theory has an internal, global shift symmetry \cite{Creminelli:2006xe,Cheung:2007st}.
This means that generic models of inflation will produce boundary correlators that are scale invariant, but not conformally invariant.
For this reason, we will develop bootstrap methods for theories that break dS boost symmetries and can therefore produce significant non-Gaussianities.

We will specifically study how to bootstrap the dS wavefunction in perturbation theory.
Given the wavefunction, one can compute late-time correlation functions by performing an additional path integral \cite{Maldacena:2002vr}.
We will leave this step implicit in our work and instead focus on the wavefunction coefficients themselves.
We first show that the momentum space discontinuity of a tree-level, $n$-point wavefunction coefficient is fixed in terms of a product of lower-point coefficients.
Equivalently, the momentum space discontinuity of a tree-level, dS Feynman diagram is equal to a product of on-shell sub-diagrams.
We then explain how to reconstruct the original $n$-point function from its discontinuity, up to contact diagram ambiguities, by using a momentum space dispersion formula.
We find that the dispersion formula for massless scalars can generically be rewritten as a contour integral of a rational function.
The dispersion formula therefore reduces complicated, nested bulk Feynman integrals to a finite sum which is completely fixed in terms of lower-point data.
Together, the dS cutting rules and dispersion formulas yield an efficient, on-shell method to compute tree-level wavefunction coefficients.

The unitarity methods used in this work are related by analytic continuation to analogous unitarity methods in AdS momentum space \cite{Meltzer:2020qbr,Meltzer:2021bmb}.\footnote{For previous works on AdS/CFT unitarity methods, from both a bulk and boundary perspective, see \cite{Fitzpatrick:2011dm,Aharony:2016dwx,Ponomarev:2019ofr,Meltzer:2019nbs}.}
The existence of an analytic continuation between the dS wavefunction and the AdS partition function \cite{Maldacena:2002vr,Harlow:2011ke} helps explain why dS wavefunction coefficients have simple analyticity properties in perturbation theory.
We should emphasize though that none of the dS identities we use rely on assuming that the AdS theory is unitary. 
For example, the factorization conditions and dispersion formulas we study are valid for heavy fields in dS, which are mapped by analytic continuation to tachyons in AdS.\footnote{Here by ``tachyons" we mean bulk AdS fields which are inconsistent with the boundary CFT unitarity bounds \cite{Simmons-Duffin:2016gjk}.}
Instead, the unitarity methods studied here follow directly from the analytic structure of tree-level dS Feynman diagrams.

Our results are closely related to and motivated by recent progress on dS unitarity methods \cite{Goodhew:2020hob,Jazayeri:2021fvk,Melville:2021lst,Goodhew:2021oqg,Baumann:2021fxj}.
It will then be useful to explain how our methods overlap with and differ from previous works on this subject.
First, here we study discontinuities of wavefunction coefficients with respect to a single Mandelstam invariant.
In \cite{Goodhew:2020hob,Melville:2021lst,Goodhew:2021oqg} a different discontinuity was studied which involves analytically continuing the norms of the external momenta, $|k_a|$.
Although these two discontinuities involve analytic continuations in different variables, they are equivalent at tree level and only start to differ at one loop.
In addition, in both this work and in \cite{Jazayeri:2021fvk,Baumann:2021fxj} tree-level exchange diagrams are computed by gluing together boundary three-point functions.
Here we perform this gluing by using a dispersion formula, where we disperse in a single Mandelstam invariant.
The output of the dispersion formula differs from the Feynman diagram result by at most a finite sum of contact diagrams.
In \cite{Jazayeri:2021fvk,Baumann:2021fxj} different gluings were performed using partial energy recursion relations.
In general, the output of these recursion relations cannot be expressed as a linear combination of dS Feynman diagrams.
Rather, the partial energy recursion relations fix part of the wavefunction coefficient and the remaining terms are fixed by imposing physical consistency conditions, including the Cosmological Optical theorem \cite{Goodhew:2020hob}, the manifest locality test \cite{Jazayeri:2021fvk}, and/or the absence of folded singularities \cite{Baumann:2021fxj}.
However, the partial energy recursion relations allow one to access certain non-analyticites which are not picked up by the dispersion formula studied here.
Overall, the results presented here and in \cite{Goodhew:2020hob,Jazayeri:2021fvk,Melville:2021lst,Goodhew:2021oqg,Baumann:2021fxj} are complementary and highlight different analyticity properties of the dS wavefunction.

\subsection*{Outline}
Here we will give an outline for the remainder of the paper.
In Section \ref{sec:dSWavefunction} we review the definition of the dS wavefunction and the Feynman rules used to compute wavefunction coefficients.
We first show that the discontinuity of the dS bulk-to-bulk propagator factorizes in terms of a product of bulk-to-boundary propagators \eqref{eq:discGgen}. 
We then show that the bulk-to-bulk propagator obeys a momentum-space dispersion formula \eqref{eq:bulkdispersion}.
From these two identities we derive the main results: a factorization condition \eqref{eq:disc_gen_4pt} and dispersion formula \eqref{eq:ambiguities} for tree-level dS wavefunction coefficients.
These results are valid for arbitrary interactions and for fields of arbitrary mass.
In Section \ref{sec:dSExchange} we apply these identities to diagrams involving massless scalars and boost-breaking interactions.
We compute general, higher-derivative, four-point exchange diagrams in the EFT of inflation \cite{Creminelli:2006xe,Cheung:2007st,Weinberg:2008hq} and also a five-point diagram constructed from a $\f'^3$ vertex.
In all cases our results are consistent with previous results in the literature.
In Section \ref{sec:AdSPerspective} we explain how the dS identities used in Sections \ref{sec:dSWavefunction} and \ref{sec:dSExchange} are related by analytic continuation to analogous AdS identities.
Finally, in Section \ref{sec:Discussion} we discuss future directions.

The appendices contain various technical details. In Appendix \ref{app:Four_Point} we give the explicit form of four-point dS exchange and contact diagrams. 
In Appendix \ref{sec:Reg_Dispersion} we discuss two different ways of regularizing divergent dispersion integrals, either by regulating the integrand with a hard cutoff or by defining the dispersion integral via analytic continuation.
We show that the two regularizations differ by a finite sum of contact diagrams. 
In Appendix \ref{app:dispersion} we study ambiguities of the dispersion formula in general asymptotically AdS spacetimes.

\subsection*{Conventions}
We use the mostly plus metric for both Lorentzian AdS$_{d+1}$ and dS$_{d+1}$.
We use bold letters for the $d$-dimensional momenta, $\mathbf{k}$, and define their norms as $k\equiv\sqrt{\mathbf{k}^{2}}$. 
We will mostly restrict to $d=3$, although our methods carry over to general spacetime dimensions.
When including indices, we use Greek letters for Lorentzian vectors, $k^{\mu}$, and Latin letters for Euclidean vectors, $k^{i}$.
Our conventions for the Fourier transform are,
\begin{align}
f(\mathbf{x})=\int \frac{d^{d}k}{(2\pi)^d} e^{\pm i\mathbf{k}\cdot \mathbf{x}}f(\mathbf{k}),
\end{align}
where we use the $+$ and $-$ signs in dS and AdS, respectively.

Since the boundary of AdS is Lorentzian while the boundary of dS is Euclidean, different sets of variables are natural to use when studying the respective boundary correlators.
We will mostly be studying the dS wavefunction, so we define the Mandelstam invariants as,
\begin{align}
s_{ab}=(\mathbf{k}_a+\mathbf{k}_b)^2.\label{eq:mandelstams}
\end{align}
In \eqref{eq:mandelstams}, $a$ and $b$ refer to different external operators.
In dS the variables $k_a$ and $\sqrt{s_{ab}}$ are often referred to as the external and internal energy variables.\footnote{In the dS literature, see for example \cite{Arkani-Hamed:2018kmz,Goodhew:2020hob}, they define $s^{\text{there}}_{ab}=\sqrt{s_{ab}^{\text{here}}}$.}
In AdS we will use the following notation,
\begin{align}
s_{-,ab}=-(\mathbf{k}_a+\mathbf{k}_b)^2,\label{eq:AdSmandelstams}
\end{align}
to emphasize the change in sign when defining the Mandelstam invariants.

At four points we define $s\equiv s_{12}$, $t\equiv s_{23}$, and $u\equiv s_{13}$.
In general QFTs, four-point functions depend on 6 independent variables, which we take to be $\{k_a,s,t\}$. 
The last Mandelstam invariant $u$ is fixed by momentum conservation.
When studying $s$-channel discontinuities, we will analytically continue in $s$ while keeping $k_a$ and $t$ held fixed.
We adopt the same conventions in AdS, except with an extra ``-" subscript for each Mandelstam invariant.

\section{dS Wavefunction}
\label{sec:dSWavefunction}

\subsection{Feynman Rules}
\label{sec:dSFeynman}
In this section we review the Feynman rules for the dS wavefunction.
We study a single scalar field $\Phi(\mathbf{x},\eta)$ in a dS$_4$ background,
\begin{align}
ds^{2}_{\text{dS}}=\frac{1}{H\eta^2}(-d\eta^2+\delta_{ij}dx^{i}dx^{j}).
\end{align}
Here $\eta\in(-\infty,\eta_0)$ is the conformal time and $\eta_0$ is the late-time cutoff where we evaluate the boundary correlators.
For simplicity, we will set the Hubble parameter to one, $H=1$.

The wavefunction of the universe is defined by taking the overlap of a field configuration at $\eta=\eta_0$ with the initial vacuum state, which we take to be the Bunch-Davies vacuum $|\Omega\>$,
\begin{align}
\Psi[\phi(\mathbf{x})]\equiv\<\phi(\mathbf{x})|\Omega\>=\int\limits^{\Phi(\mathbf{x},\eta_0)=\phi(\mathbf{x}) }_{ \Phi(\mathbf{x},-\infty)=0  }\mathcal{D}\Phi \hspace{.05cm} e^{i S[\Phi]}.\label{eq:wavefunction_def}
\end{align}
In weakly-coupled theories we can expand the wavefunction in terms of the late-time fluctuations,
\begin{align}
\Psi[\phi]\approx \text{exp}\left[-\sum\limits_{n=2}^{\infty}\frac{1}{n!}\frac{d^{3}k_1\ldots d^{3}k_n}{(2\pi)^{3n}}\Psi_{n}(\mathbf{k}_1,\ldots,\mathbf{k}_n)\phi(\mathbf{k}_1)\ldots\phi(\mathbf{k}_n)\right].
\end{align}
The functions $\Psi_{n}$ are known as the wavefunction coefficients. 
Given the wavefunction, we can compute the late-time correlators using \cite{Maldacena:2002vr},
\begin{align}
\<\f(\mathbf{k}_1)\ldots \f(\mathbf{k}_n)\>=\frac{\int \mathcal{D}\phi \hspace{.05cm} \phi(\mathbf{k}_1)\ldots \phi(\mathbf{k}_n)|\Psi[\phi]|^2}{\int \mathcal{D}\phi |\Psi[\phi]|^2}\label{eq:wavefunction_to_correlator}.
\end{align}
At tree level it is straightforward to go from the wavefunction coefficients to the late-time correlators.
However, in this work we will only compute the wavefunction itself.
Finally, it will be convenient to remove an overall momentum conserving $\delta$-function from the wavefunction coefficients,
\begin{align}
\Psi_{n}(\mathbf{k}_1,\ldots,\mathbf{k}_n)=(2\pi)^3\delta(\mathbf{k}_1+\ldots+\mathbf{k}_n)\psi_n(\mathbf{k}_1,\ldots,\mathbf{k}_n).
\end{align}

The wavefunction coefficients can be computed using a set of Feynman rules that are related to the ones used in AdS/CFT \cite{Maldacena:2002vr} (see also \cite{Anninos:2014lwa,Goon:2018fyu}). 
To start, we write down the free action for a field $\Phi$ propagating on a dS$_4$ background with a speed of sound $c_s$,
\begin{align}
S^{\text{dS}}_{0}=\frac{1}{2}\int d\eta d^3\mathbf{x}\frac{1}{\eta^2}\left(\Phi'^2-c_s^2(\partial_i\Phi)^2-\frac{m^{2}}{\eta^2}\Phi^2\right). \label{eq:freeactiondS}
\end{align}
Here the prime denotes a derivative with respect to $\eta$.
A massless, self-interacting scalar with a non-trivial speed of sound appears in the decoupling limit of the EFT of inflation \cite{Creminelli:2006xe,Cheung:2007st}.
In that context, $\Phi$ is the Goldstone boson for broken time translations and the decoupling limit is,
\begin{align}
M_{\text{pl}}\rightarrow\infty \ \text{ and } \ \dot{H}\rightarrow0 \ \text{ such that } \ M_{\text{pl}}^{2}\dot{H}\text{ is fixed.}
\end{align}
Here $M_{\text{pl}}$ is the Planck scale and the dot denotes a derivative with respect to the physical time $t$.
In this limit gravity is decoupled and the EFT of inflation becomes a non-relativistic, scalar QFT on a rigid background, which we have taken to be dS.
For more details on the EFT of inflation and its decoupling limit see \cite{Creminelli:2006xe,Cheung:2007st,Baumann:2011su,Piazza:2013coa,Baumann:2014nda}.

A non-trivial, constant speed of sound, $c_s\neq1$ preserves boundary scale invariance, but breaks boundary special conformal transformations \cite{Baumann:2015xxa}.\footnote{A time-dependent speed of sound breaks both scale and special conformal transformations on the boundary.}
Scale invariance is also broken during inflation, but its breaking is slow-roll suppressed while the breaking of special conformal transformations is generically of order one \cite{Green:2020ebl}.
We will ignore slow-roll corrections and assume that the boundary correlators are scale, but not conformally, invariant.

Since we are studying a single bulk scalar field, we can in practice set $c_s=1$ and restore the $c_s$ dependence later using dimensional analysis. Equivalently, we are rescaling the coordinates to impose a fake dS invariance.
Setting $c_s=1$, the action \eqref{eq:freeactiondS} yields the standard relativistic dS propagators,
\begin{align}
\mathcal{K}_{\nu}(k,\eta,\eta_0)=&\left(\frac{\eta}{\eta_0}\right)^{\frac{3}{2}}\frac{H_{\nu}^{(2)}(-k\eta)}{H^{(2)}_{\nu}(-k\eta_0)},\label{eq:K_dS}
\\[5pt]
\mathcal{G}_\nu(k,\eta_1,\eta_2,\eta_0)=&\frac{\pi}{4}\bigg(H^{(1)}_{\nu}(-k\eta_1)H^{(2)}_{\nu}(-k\eta_2)\theta(\eta_1-\eta_2)+H^{(1)}_{\nu}(-k\eta_2)H^{(2)}_{\nu}(-k\eta_1)\theta(\eta_2-\eta_1)
\nonumber \\
&\hspace{.6cm}-\frac{H^{(1)}_{\nu}(-k\eta_0)}{H^{(2)}(-k\eta_0)}H^{(2)}_{\nu}(-k\eta_1)H^{(2)}_{\nu}(-k\eta_2)\bigg),\label{eq:G_dS}
\end{align}
where $\mathcal{K}_{\nu}$ is the bulk-to-boundary propagator and $\mathcal{G}$ is the bulk-to-bulk propagator.
The $H^{(i)}_{\nu}$ are the Hankel functions and the parameter $\nu$ is related to the mass $m$ of the bulk field $\Phi$ by,
\begin{align}
\nu=\sqrt{\frac{9}{4}-m^2}.\label{eq:mass_to_dim_dS}
\end{align}
Requiring that we have a Bunch-Davies vacuum in the infinite past translates into the following boundary conditions, 
\begin{align}
\lim\limits_{\eta\rightarrow-\infty} \mathcal{K}_{\nu}(k,\eta,\eta_0)=0,\quad
\lim\limits_{\eta_1\rightarrow-\infty} \mathcal{G}_{\nu}(k,\eta_1,\eta_2,\eta_0)=0,\label{eq:BD_bdy_conditions}
\end{align}
which hold if we use the $i\epsilon$ prescription, $k\rightarrow k-i\epsilon$.
We will often leave the dependence on $\eta_0$ implicit and take the limit $\eta_0\rightarrow 0$ for IR-finite diagrams.

Next, we can consider interactions of $\Phi$ which preserve boundary scale and Poincar\'e invariance, but break special conformal transformations. The most general such interaction takes the form \cite{Jazayeri:2021fvk},
\begin{align}
S_{\text{int}}=\int d\eta d^{3}\mathbf{x}\ \eta^{\sum_{a}N_a-4}\partial^{N_1}\Phi\ldots\partial^{N_n}\Phi,\label{eq:gen_boost_breaking_action}
\end{align} 
where $\partial$ denotes a derivative with respect to $\eta$ or $x^i$ and $N_a$ counts the total number of derivatives, of either type, acting on each field.
The powers of $\eta$ are fixed by imposing boundary scale invariance.
These types of interactions appear in the EFT of inflation and we will study explicit examples of them in Section \ref{sec:dSExchange}.

Finally, to compute wavefunction coefficients we draw dS Feynman diagrams where all external lines end on the future boundary at $\eta=\eta_0$.
We use $\mathcal{K}_{\nu}$ and $\mathcal{G}_{\nu}$ for each bulk-to-boundary and bulk-to-bulk line, respectively, and include a factor of $iV$ for each interaction vertex.
We then integrate over all loop momenta and the positions in time of the interaction vertices.
We will mostly study $s$-channel, exchange diagrams of the form,
\begin{align}
	\psi_{\text{exch}}(k_a,s,t)=\begin{aligned}
		\includegraphics[scale=.3]{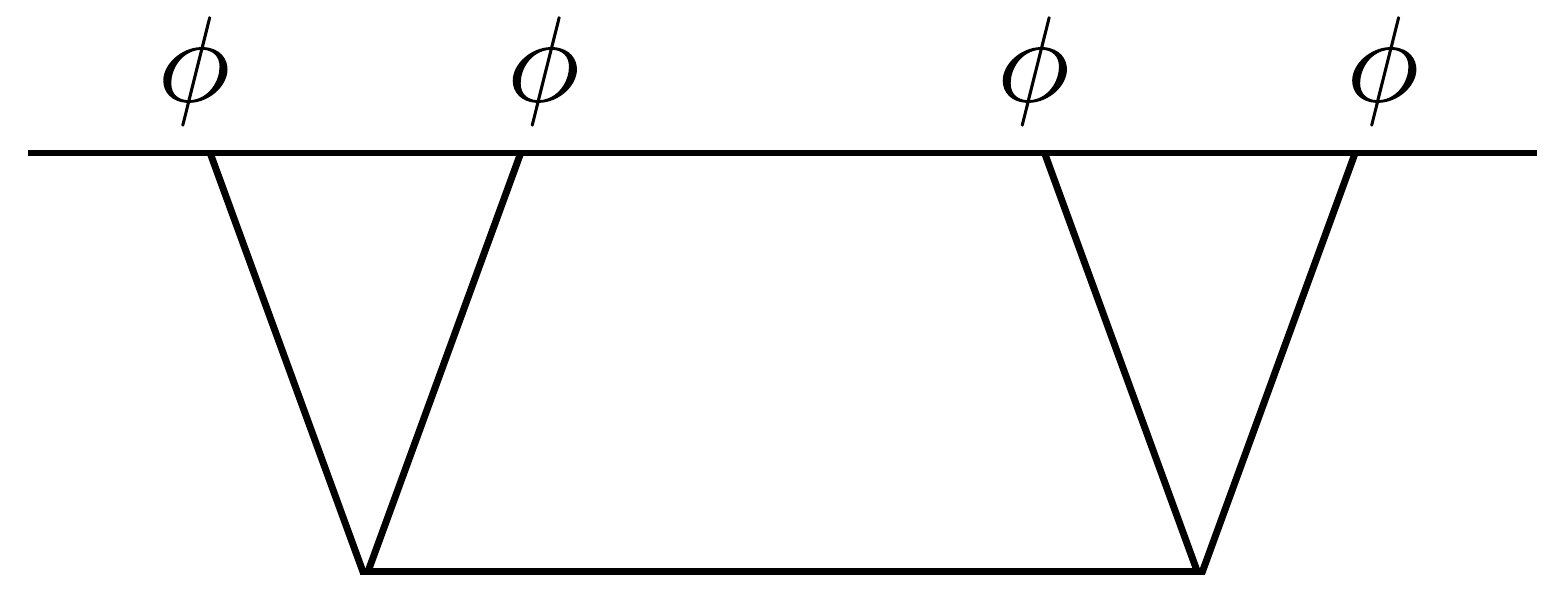}
	\end{aligned}.
\end{align}

\subsection{Cutting Rules and Dispersion}
In this section we will show that tree-level dS wavefunction coefficients can be computed using cutting rules and momentum space dispersion formulas.
For simplicity, we start by studying the propagators of a conformally coupled scalar, which corresponds to $m^{2}=2$ in Hubble units or $\nu=1/2$.
We find,
\begin{align}
\mathcal{K}_{\nu=1/2}(k,\eta)&=\frac{\eta}{\eta_0}e^{ik(\eta-\eta_0)},\label{eq:dSKconf}
\\
\mathcal{G}_{\nu=1/2}(k,\eta_1,\eta_2)&=\frac{\eta_1\eta_2}{2k}\bigg(\theta (\eta_1-\eta_2) e^{-i k (\eta_1-\eta_2)}+\theta (\eta_2-\eta_1) e^{i k (\eta_1-\eta_2)}-e^{-i k (2 \eta_0-\eta_1-\eta_2)}\bigg).\label{eq:dSGconf}
\end{align}
As stated earlier, to ensure that both propagators decay at past infinity, we give the norms $k$ a negative imaginary part.
In polar coordinates, $k^{2}=re^{i\theta}$, this becomes the condition $\theta\in(-2\pi,0)$.
If $\theta$ is in this interval, then the Feynman integrals converge.
It is therefore natural to place the $k^2$ branch cut on the positive real axis. 
We then define the following monodromy operation,
\begin{align}
\text{disc}_{k^{2}}f(k^2)=f(e^{-2\pi i}k^2)-f(k^2),
\end{align}
which in our conventions corresponds to a discontinuity when $k^2$ is real and positive.
For the conformally coupled scalar  we find,
\begin{align}
\text{disc}_{k^{2}}\mathcal{G}_{1/2}(k,\eta_1,\eta_2)=P_{1/2}(k)&\text{disc}_{k^{2}}\mathcal{K}_{1/2}(k,\eta_1)\text{disc}_{k^{2}}\mathcal{K}_{1/2}(k,\eta_2),\label{eq:discGconf}
\end{align}
where the power spectrum $P_{\nu}(k)$ is defined for general $\nu$ by,
\begin{align}
\<\phi(\mathbf{k},\eta_0)\phi(\mathbf{k'},\eta_0)\>=\delta(\mathbf{k}+\mathbf{k'})P_{\nu}(k),
\\[3pt]
P_{\nu}(k)=\frac{\pi}{4}\eta_0^3H^{(1)}_{\nu}(-k\eta_0)H^{(2)}_{\nu}(-k\eta_0).
\end{align}
The relation \eqref{eq:discGconf} shows that $\disc_{k^2}$ of the dS bulk-to-bulk propagator produces a result which is factorized in $(\eta_1,\eta_2)$.
It is not difficult to show that this identity continues to hold for generic $\nu$,
\begin{align}
\text{disc}_{k^{2}}\mathcal{G}_{\nu}(k,\eta_1,\eta_2)=P_{\nu}(k)&\text{disc}_{k^{2}}\mathcal{K}_{\nu}(k,\eta_1)\text{disc}_{k^{2}}\mathcal{K}_{\nu}(k,\eta_2).\label{eq:discGgen}
\end{align}
The identity \eqref{eq:discGgen} can be represented diagrammatically as,
\begin{align}
	\begin{aligned}
		\includegraphics[scale=.275]{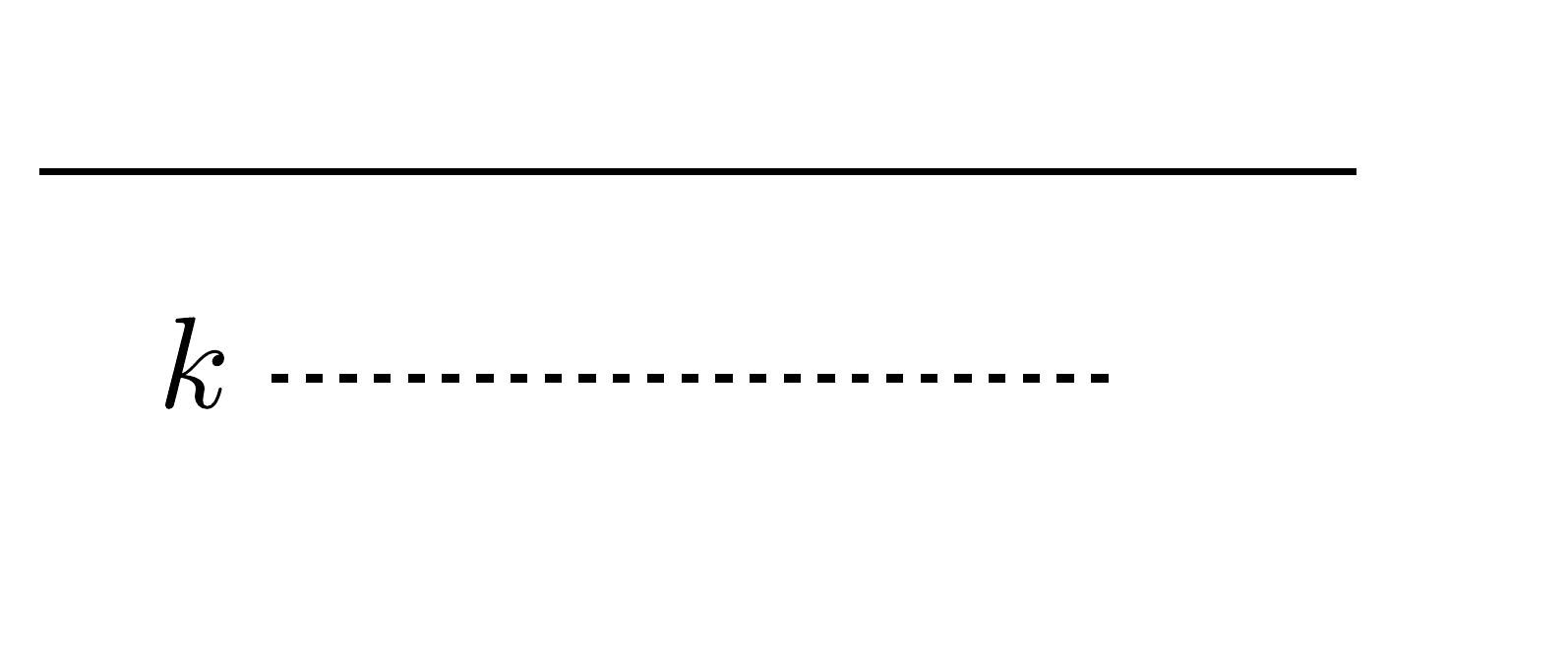}
	\end{aligned}
	= \ P_{\nu}(k) \ \begin{aligned}
		\includegraphics[scale=.275]{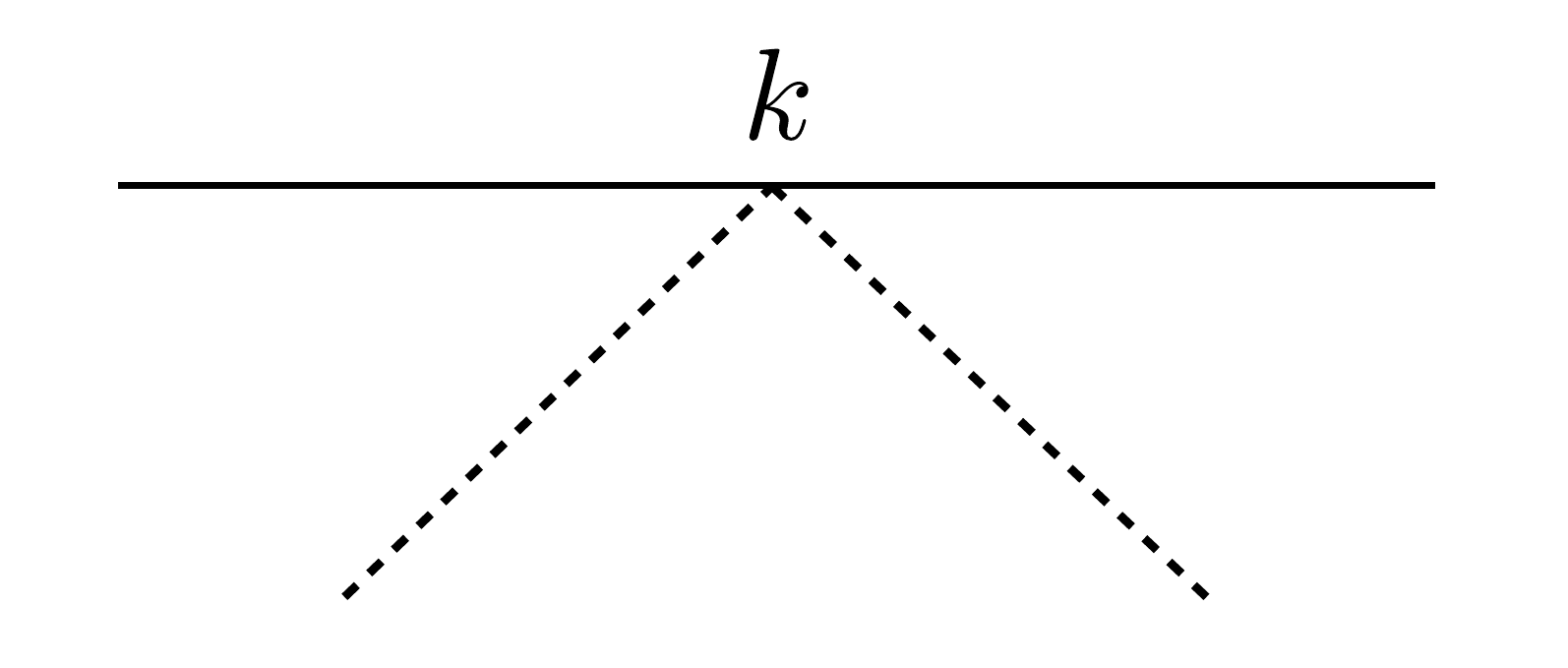}
	\end{aligned},
\end{align}
where the dashed lines correspond to the discontinuity of the appropriate propagator.
The fact that the discontinuity of the bulk-to-bulk propagator factorizes has been observed previously in both AdS \cite{Meltzer:2020qbr} and dS \cite{Goodhew:2021oqg,Baumann:2021fxj}.
To prove \eqref{eq:discGgen} we used \cite{Hankel}:
\begin{align}
\sin(\nu\pi)H^{(1)}_{\nu}(ze^{i\pi m})&=-\sin((m-1)\nu \pi)H^{(1)}_{\nu}(z)-e^{-i\pi\nu}\sin(m\nu\pi)H^{(2)}_{\nu}(z),
\\
\sin(\nu\pi)H^{(2)}_{\nu}(ze^{i\pi m})&=e^{i\pi\nu}\sin(m\nu\pi)H^{(1)}_{\nu}(z)+\sin((m+1)\nu\pi)H^{(2)}_{\nu}(z),
\end{align}
where $m\in\mathbb{Z}$ and $\nu$ is arbitrary.

To see why the factorization property \eqref{eq:discGgen} is useful, we will study the discontinuity of a tree-level exchange diagram.
We will consider an exchange diagram constructed from a non-derivative $\phi^2\xi$ interaction,\footnote{To simplify notation, we will use the same letter for the bulk field and its boundary condition at $\eta_0$.} where $\xi$ is a generic scalar:
\begin{align}
\centering
\psi_{\xi}(k_a,s,t)&=\begin{aligned}
		\includegraphics[scale=.275]{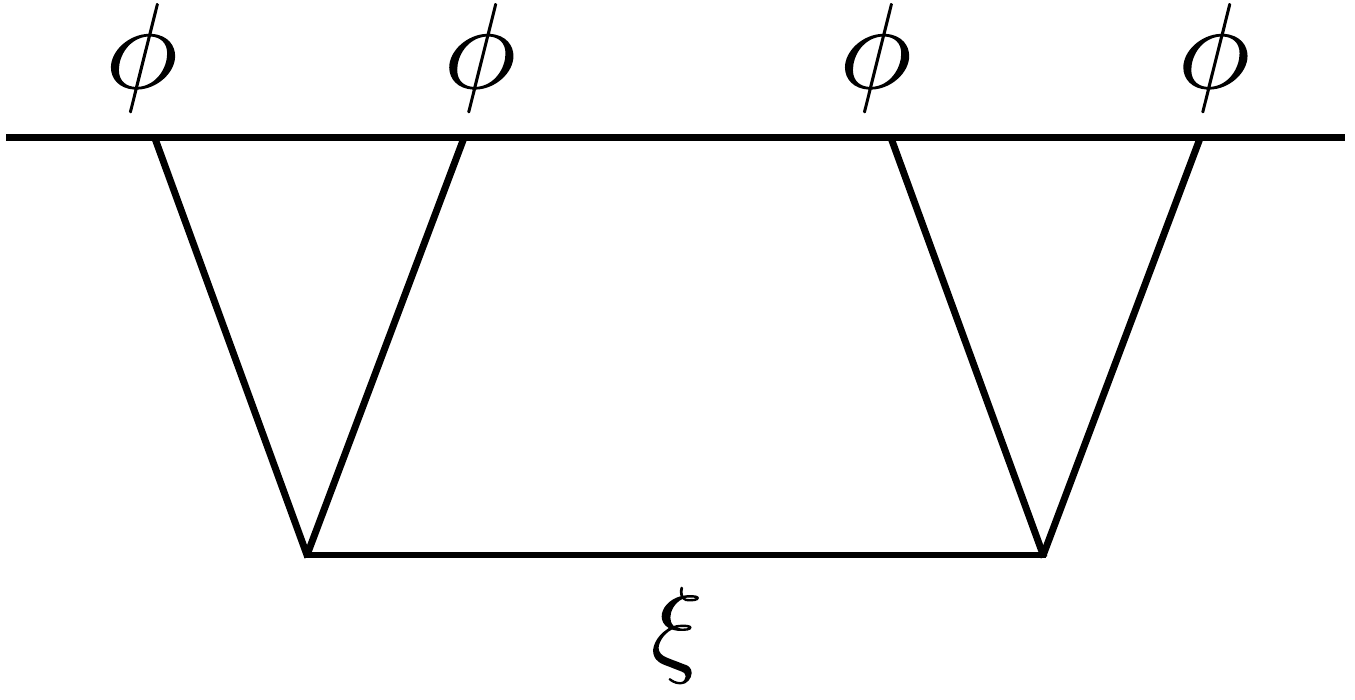}
	\end{aligned}
\nonumber
\\
&=-g^2\int\limits_{-\infty}^{\eta_0} \frac{d\eta_1d\eta_2}{\eta_1^4\eta_2^4}\mathcal{K}_{\f}(k_1,\eta_1)\mathcal{K}_{\f}(k_2,\eta_1)
\mathcal{G}_{\xi}(\sqrt{s},\eta_1,\eta_2)\mathcal{K}_{\f}(k_3,\eta_2)\mathcal{K}_{\f}(k_4,\eta_2).\label{eq:exchangepsixi}
\end{align}
Here we used the fields to label the propagators and, as a reminder, have defined $s=(\mathbf{k}_1+\mathbf{k}_2)^2$.
From the factorization property of the bulk-to-bulk propagator \eqref{eq:discGgen}, we see that taking the discontinuity of \eqref{eq:exchangepsixi} factorizes the diagram into a product of three-point functions,
\begin{align}
\text{disc}_{s}\psi_{\xi}(k_a,s,t)=P_{\xi}(\sqrt{s})\text{disc}_{s}\psi_{3}^{\f\f\xi}(k_1,k_2,\sqrt{s})\text{disc}_{s}\psi_{3}^{\f\f\xi}(k_3,k_4,\sqrt{s}),\label{eq:cut_xi_exchange}
\end{align}
where the three-point wavefunction coefficient is,
\begin{align}
\psi_{3}^{\f\f\xi}(k_1,k_2,k_3)=ig\int\limits_{-\infty}^{\eta_0}\frac{d\eta}{\eta^4} \mathcal{K}_{\f}(k_1,\eta)\mathcal{K}_{\f}(k_2,\eta)\mathcal{K}_{\xi}(k_3,\eta).
\end{align}
In terms of diagrams, the factorization condition \eqref{eq:cut_xi_exchange} is,
\begin{align}
\disc_{s}\hspace{.1cm}\begin{aligned}
		\includegraphics[scale=.275]{xi_exchange.pdf}
	\end{aligned}=P_{\xi}(\sqrt{s})\hspace{.1cm}\begin{aligned}
		\includegraphics[scale=.275]{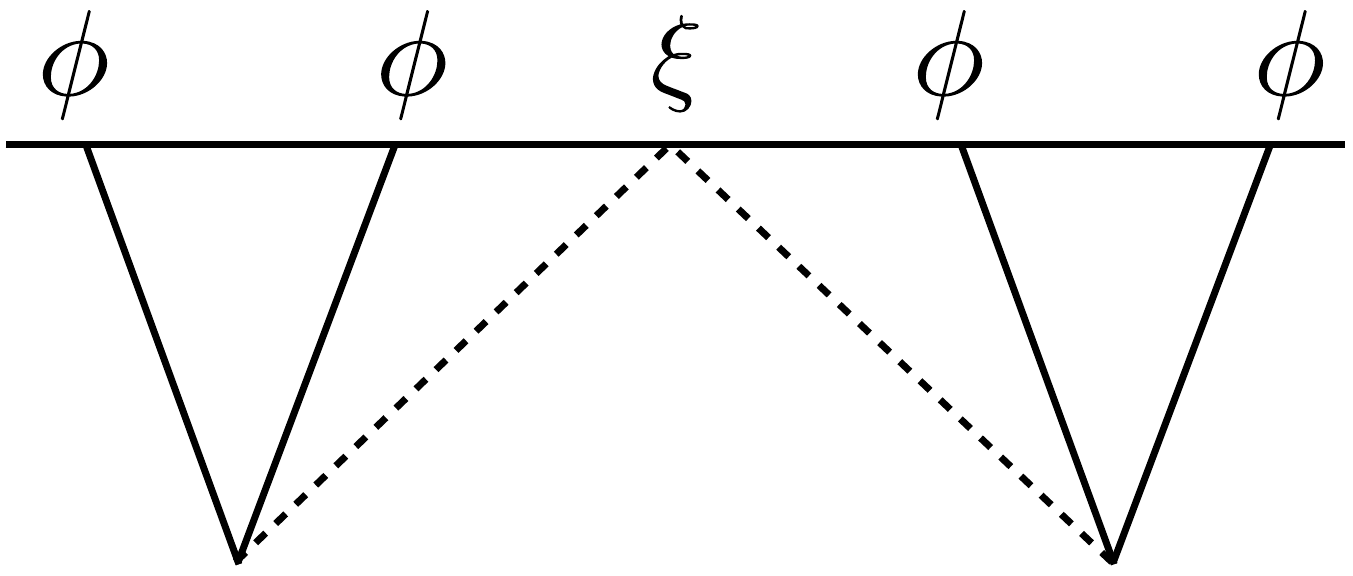}
	\end{aligned}.\label{fig:cut_diagram}
\end{align}

To compute the full diagram, we will use that $\mathcal{G}_{\nu}$ is an analytic function of $k^2$ that decays at large $k$.
This allows us to write down a dispersion formula for the bulk-to-bulk propagator,
\begin{align}
\mathcal{G}_{\nu}(k,\eta_1,\eta_2)=\frac{1}{2\pi i}\int\limits_{0}^{\infty}\frac{dp^2}{p^2-k^2+i\epsilon}\text{disc}_{p^{2}}\mathcal{G}_{\nu}(p,\eta_1,\eta_2).\label{eq:bulkdispersion}
\end{align}
The relation \eqref{eq:bulkdispersion} makes it manifest that all tree-level dS diagrams involving scalar exchange have a dispersive representation, up to possible contact diagram ambiguities.
For example, if we use \eqref{eq:bulkdispersion} in \eqref{eq:exchangepsixi} we find the following fixed-$t$ dispersion formula for the $\xi$ exchange diagram,
\begin{align}
\psi_{\xi}(k_a,s,t)&=-g^2\frac{1}{2\pi i}\int\limits_{0}^{\infty}\frac{ds'}{s'-s+i\epsilon}\int\limits_{-\infty}^{\eta_0} \frac{d\eta_1d\eta_2}{\eta_1^4\eta_2^4}\mathcal{K}_{\f}(k_1,\eta_1)\mathcal{K}_{\f}(k_2,\eta_1)
\text{disc}_{s'}\mathcal{G}_{\xi}(\sqrt{s'},\eta_1,\eta_2)
\nonumber 
\\
&\hspace{2.157in}\times\mathcal{K}_{\f}(k_3,\eta_2)\mathcal{K}_{\f}(k_4,\eta_2)
\nonumber \\
&=\frac{1}{2\pi i}\int\limits_0^{\infty}\frac{ds'}{s'-s+i\epsilon}\text{disc}_{s'}\psi_{\xi}(k_a,s',t).\label{eq:disp_simple_dS}
\end{align}
The identity \eqref{eq:disp_simple_dS} is true for any field $\xi$, including both light $(m_{\xi}^{2}<9/4)$ and heavy fields $(m_{\xi}^{2}>9/4)$, because the bulk identity \eqref{eq:bulkdispersion} is true for all $\nu$.
In addition, \eqref{eq:disp_simple_dS} is also manifestly true for IR-divergent diagrams because we have cut off the $\eta$-integral.
The observation that a dispersion formula for the bulk-to-bulk propagator implies one for the boundary correlator has also been used in AdS \cite{Meltzer:2021bmb}.

To be more general, we can consider the exchange of a scalar $\chi$ which couples to $\phi$ through derivative interactions,
\begin{align}
\psi_{\chi}(k_a,s,t)&=-\int\limits_{-\infty}^{\eta_0} \frac{d\eta_1d\eta_2}{\eta_1^4\eta_2^4}V_LV_R\mathcal{K}_{\f}(k_1,\eta_1)\mathcal{K}_{\f}(k_2,\eta_1)
\mathcal{G}_{\chi}(\sqrt{s},\eta_1,\eta_2)\mathcal{K}_{\f}(k_3,\eta_2)\mathcal{K}_{\f}(k_4,\eta_2),\label{eq:exchange_general}
\end{align}
where $V_{L,R}$ are the left and right vertex factors. 
At tree level these are polynomials in the external momenta, $\mathbf{k}_a\cdot\mathbf{k}_b$, and differential operators in $\eta_{1,2}$.
Taking the discontinuity in $s$ of \eqref{eq:exchange_general} again factorizes the diagram in terms of three-point functions,
\begin{align}
\text{disc}_{s}\psi_{\chi}(k_a,s,t)=P_{\chi}(\sqrt{s})\text{disc}_{s}\psi_{3}^{\f\f\chi}(k_1,k_2,\sqrt{s})\text{disc}_{s}\psi_{3}^{\f\f\chi}(k_3,k_4,\sqrt{s}),
\label{eq:disc_gen_4pt}
\end{align}
and therefore the bulk picture \eqref{fig:cut_diagram} is still valid.
However, the presence of the non-trivial vertex factors, and specifically the dependence on the dot products $\mathbf{k}_a\cdot\mathbf{k}_b$, means that the previous dispersion formula \eqref{eq:disp_simple_dS} will in general only hold up to contact diagram ambiguities,
\begin{align}
\psi_{\chi}(k_a,s,t)=\frac{1}{2\pi i}\int\limits_0^{\infty}\frac{ds'}{s'-s+i\epsilon}\text{disc}_{s'}\psi_{\chi}(k_a,s',t)+\text{(contact diagrams)}.\label{eq:ambiguities}
\end{align}
The discontinuity of a contact diagram with respect to a Mandelstam invariant manifestly vanishes,
\begin{align}
\disc_{s,t,u}\psi_{\text{cont}}(k_a,s,t)=0,
\end{align}
so the relation \eqref{eq:ambiguities} is consistent with the factorization condition \eqref{eq:disc_gen_4pt}.
In section \ref{sec:AdSPerspective} we will prove \eqref{eq:ambiguities} in AdS for general, boost-breaking interactions.
The corresponding result in dS for light field exchange then follows by analytic continuation (for an argument valid for heavy fields in dS see Appendix \ref{app:dispersion}).

In the next section we will study the dispersion formula \eqref{eq:ambiguities} for exchange diagrams involving boost-breaking, derivative interactions.
We will see that contact diagram ambiguities only appear for interactions involving spatial derivatives, such as $\f'(\partial_i\f)^2$.
If the diagram is constructed using interactions that only involve time derivatives, then the dispersion formula \eqref{eq:ambiguities} holds without any ambiguities.
There is a similar story when we have internal or external spinning fields: taking the discontinuity still factorizes the four-point function, but there can be contact diagram ambiguities in the dispersion formula due to derivative interactions.
The only new feature for spinning fields is that we need to include a sum over helicities in \eqref{eq:disc_gen_4pt}.

One of the goals of the cosmological bootstrap is to develop a purely boundary way to study and think about inflationary correlators.
It may look like we have taken a step backwards in this regard since the factorization condition \eqref{eq:disc_gen_4pt} and boundary dispersion formula \eqref{eq:ambiguities} were derived by using explicit properties of the dS bulk-to-bulk propagator.
While this is true, these identities are useful because they give relations between purely boundary, or on-shell, observables. 
Therefore, we can use \eqref{eq:disc_gen_4pt} and \eqref{eq:ambiguities} to study four-point functions by gluing together boundary three-point functions, which in turn can be bootstrapped using symmetries and consistency conditions \cite{Pajer:2020wxk}.

\section{Boost-Breaking dS Diagrams}
\label{sec:dSExchange}
In this section we study boost-breaking dS exchange diagrams that appear in the EFT of inflation.
We will not consider boost-preserving cubic interactions as this analysis was previously carried out in AdS \cite{Meltzer:2020qbr,Meltzer:2021bmb} and the corresponding dS analysis is identical in form.

For the remainder of this section we will use $\phi$ to denote a massless ($m=0$) scalar in dS.
It is dual to a boundary scalar with dimension $\Delta_\f=3$, or $\nu_\f=3/2$.
When studying three-point functions it will be convenient to introduce the following symmetric polynomials,
\begin{align}
e_1=k_1+k_2+k_3,\quad e_2=k_1k_2+k_1k_3+k_2k_3, \quad e_3=k_1k_2k_3.
\end{align}
The first polynomial, $e_1$, is also the total energy variable, $k_T=\sum_ak_a$, at three points.
\subsection{$\f\f'^2$}
We first study the following cubic interaction,
\begin{align}
S_{\text{int}}=\frac{g}{2}\int d^3xd\eta\frac{1}{\eta^2}\phi\phi'^2.
\end{align}
We can prove that the exchange diagram constructed from this interaction obeys a fixed-$t$ dispersion formula without any contact diagram ambiguities.
The exchange diagram is,
\begin{align}
\psi^{\f\f'^2}_{4}(k_a,s,t)=-g^2\int\limits_{-\infty}^{0} d\eta_1d\eta_2& \ \mathcal{G}(\sqrt{s},\eta_1,\eta_2)F^{\f\f'^2}_3(k_1,k_2,\eta_1)F^{\f\f'^2}_3(k_3,k_4,\eta_2),
\label{eq:exchange_test}
\end{align} 
where,
\begin{align}
F^{\f\f'^2}_3(k_1,k_2,\eta_1)=\eta_1^{-2}\mathcal{K}'_\f(k_1,\eta_1)\mathcal{K}'_\f(k_2,\eta_1)-\left[\partial_{\eta_1}(\eta_1^{-2}\mathcal{K}'_\f(k_1,\eta_1)\mathcal{K}_\f(k_2,\eta_1))+(k_1\leftrightarrow k_2)\right].
\end{align}
To derive a dispersion formula for $\psi^{\f\f'^2}_{4}$ we can use the bulk dispersion formula \eqref{eq:bulkdispersion} in \eqref{eq:exchange_test} and then switch the orders of integration.\footnote{For IR-divergent interactions the integrals can only be swapped if we cut off the $\eta$-integral.} This yields,
\begin{align}
\psi^{\f\f'^2}_{4}(k_a,s,t)=\frac{1}{2\pi i}\int\limits_0^{\infty}\frac{ds'}{s'-s+i\epsilon}\text{disc}_{s'}\psi^{\f\f'^2}_{4}(k_a,s',t)\label{eq:bulkdisptest}.
\end{align}

The virtue of the dispersive representation \eqref{eq:bulkdisptest} is that the dispersion integral is simpler to compute in comparison to the bulk Feynman integrals in \eqref{eq:exchange_test}.
To compute \eqref{eq:bulkdisptest}, we first use that the $s$-channel discontinuity factorizes the exchange diagram into a product of three-point functions, \eqref{eq:disc_gen_4pt}.
The three-point wavefunction coefficient for $\f\f'^2$ is,
\begin{align}
\psi^{\f\f'^2}_{3}(k_1,k_2,k_3)=\frac{g}{e_1^2}(e_2e_3+e_1e_2^2-2e_1^2e_3).
\end{align}
Plugging this result into \eqref{eq:disc_gen_4pt} yields,
\begin{align}
\text{disc}_{s}\psi^{\f\f'^2}_{4}(k_a,s,t)=& \  P_{3/2}(\sqrt{s})\text{disc}_{s}\psi^{\f\f'^2}_{3}(k_1,k_2,\sqrt{s})\text{disc}_{s}\psi^{\f\f'^2}_{3}(k_3,k_4,\sqrt{s})
\nonumber \\[4pt]
 =& \  \frac{2 g^2 s^{3/2} }{\left((k_1+k_2)^2-s\right)^2 \left((k_3+k_4)^2-s\right)^2}
\nonumber \\
 & \ \left(k_1^4+4 k_1^3 k_2+4 k_1^2 k_2^2+4 k_1 k_2^3+k_2^4-s \left(k_1^2+k_2^2\right)\right) 
\nonumber \\
 & \ \left(k_3^4+4 k_3^3 k_4+4 k_3^2 k_4^2+4 k_3 k_4^3+k_4^4-s \left(k_3^2+k_4^2\right)\right).\label{eq:disc_phiphipsq}
\end{align}
In \eqref{eq:disc_phiphipsq} we evaluated the power spectrum in the limit $\eta_0\rightarrow0$,
\begin{align}
P_{3/2}(k)\bigg|_{\eta_0=0}=\frac{1}{2k^3}.
\end{align}

To compute the dispersion integral \eqref{eq:bulkdisptest}, we can note that \eqref{eq:disc_phiphipsq} is a rational, odd function of $\sqrt{s}$.
For this reason, we will make the change of variables $s'=p^2$,
\begin{align}
\psi^{\f\f'^2}_{4}(k_a,s,t)=& \ \frac{1}{\pi i}\int\limits_0^{\infty}\frac{dp}{p^2-s+i\epsilon}p\hspace{.05cm} \text{disc}_{p^2} \psi^{\f\f'^2}_{4}(k_a,p^2,t).
\label{eq:disp_phiphipsq_Pt2_0}
\end{align}
The integrand is now an even function of $p$, so we can extend the integration over the entire real line,
\begin{align}
\psi^{\f\f'^2}_{4}(k_a,s,t)=& \ \frac{1}{2\pi i}\int\limits_{-\infty}^{\infty}\frac{dp}{p^2-s+i\epsilon} \frac{2 g^2 p^4 }{\left((k_1+k_2)^2-p^2\right)^2 \left((k_3+k_4)^2-p^2\right)^2}
\nonumber \\
 & \hspace{1.34in} \left(k_1^4+4 k_1^3 k_2+4 k_1^2 k_2^2+4 k_1 k_2^3+k_2^4-p^2 \left(k_1^2+k_2^2\right)\right) 
\nonumber \\
 & \hspace{1.34in} \left(k_3^4+4 k_3^3 k_4+4 k_3^2 k_4^2+4 k_3 k_4^3+k_4^4-p^2 \left(k_3^2+k_4^2\right)\right).
 \label{eq:disp_phiphipsq_Pt2}
\end{align}
The integrand of \eqref{eq:disp_phiphipsq_Pt2} vanishes as $p\rightarrow\infty$ in the  complex plane, and therefore we can close the $p$-contour in either the upper or lower half-plane.
The poles in $p$ are located at,
\begin{align}
p&=\pm(k_1+k_2),
\\
p&=\pm(k_3+k_4),
\\
p&=\pm\sqrt{s-i\epsilon}.
\end{align}
When evaluating the contour integral, we should recall that the Bunch-Davies boundary condition \eqref{eq:BD_bdy_conditions} implies that the $k_a$ have a small, negative imaginary part.
Finally, we can close the contour in the lower half-plane to find,
\begin{align}
\psi^{\f\f'^2}_{4}(k_a,s,t)=-\left(\res\limits_{p=k_1+k_2}+\res\limits_{p=k_3+k_4}+\res\limits_{p=\sqrt{s-i\epsilon}}\right)\frac{p}{p^2 -s+i\epsilon}\text{disc}_{p^2}\psi^{\f\f'^2}_{4}(k_a,p^2,t).\label{eq:test_sum_result}
\end{align}
We see that the bulk Feynman integrals in \eqref{eq:exchange_test} have been reduced to a finite sum.\footnote{The same contour integrals were used in \cite{Raju:2012zr,Raju:2012zs,Albayrak:2018tam,Albayrak:2019yve} to study gauge and graviton correlators in AdS$_4$.}

Although the sum in \eqref{eq:test_sum_result} is simple to evaluate, the final result is lengthy and we leave the full result for Appendix \ref{app:Four_Point}.
Here we will only point out that the expression \eqref{eq:test_sum_result} obeys a number of consistency conditions.
For example, it does not have folded singularities, i.e. it is finite in the limits $s\rightarrow (k_1+k_2)^2$ and $s\rightarrow (k_3+k_4)^2$.
In addition, it obeys the manifest locality test (MLT) \cite{Jazayeri:2021fvk}:
\begin{align}
\partial_{k_1}\psi^{\f\f'^2}_{4}(k_a,s,t)\bigg|_{k_1=0}=0.\label{eq:MLT_test}
\end{align}
The MLT \eqref{eq:MLT_test} holds for massless scalar fields in dS whose interactions only involve non-negative powers of $\partial_{i}$ \cite{Jazayeri:2021fvk}.\footnote{As emphasized in \cite{Jazayeri:2021fvk}, there exist theories which are local, but not manifestly so. For example, if we minimally couple a scalar to gravity and integrate out the constrained fields, then we generate powers of $\partial_{i}^{-2}$.}
Of course, the fact that these consistency conditions hold is expected since we have simply rewritten a Feynman integral using a bulk dispersion formula.

\subsection{EFT of Inflation}
\label{sec:EFT_inflation_examples}
\subsubsection{Leading Order}
In this section we study the leading cubic interactions for the EFT of inflation in the decoupling limit:
\begin{align}
S_{\text{int}}=\int\limits d\eta d^3x\frac{1}{\eta}\left(\frac{g_1}{3!}\f'^3+\frac{g_2}{2}\f'(\partial_i\f)^2\right).
\end{align}

We start by studying an exchange diagram constructed from two copies of the $\f'^3$ vertex. 
As in our previous example, we can prove that the dispersion formula reconstructs this exchange diagram without ambiguities by studying the bulk Feynman diagram,
\begin{align}
\psi^{\f'^3}_{4}(k_a,s,t)=-g_1^2\int\limits_{-\infty}^{0} d\eta_1d\eta_2& \ \mathcal{G}_\f(\sqrt{s},\eta_1,\eta_2)
\partial_{\eta_1}\left(\eta_1^{-1}\mathcal{K}'_\f(k_1,\eta_1)\mathcal{K}'_\f(k_2,\eta_1)\right)
\nonumber
\\
&
\partial_{\eta_2}\left(\eta_2^{-1}\mathcal{K}'_\f(k_3,\eta_2)\mathcal{K}'_\f(k_4,\eta_2)\right).
\label{eq:EFT1sqFeynman}
\end{align}
To derive the dispersive representation of $\psi^{\f'^3}_{4}$, we use the bulk dispersion formula \eqref{eq:bulkdispersion} in \eqref{eq:EFT1sqFeynman} and then switch the orders of integration. This gives,
\begin{align}
\psi^{\f'^3}_{4}(k_a,s,t)=\frac{1}{2\pi i}\int\limits_{0}^{\infty}\frac{ds'}{s'-s+i\epsilon}\text{disc}_{s'}\psi^{\f'^3}_{4}(k_a,s',t).\label{eq:dispphipcubed}
\end{align}
To compute the above discontinuity, we need the three-point wavefunction coefficient for the $\f'^3$ interaction,
\begin{align}
\psi^{\f'^3}_{3}=2g_1\frac{e_3^2}{e_1^3}.\label{eq:EFT1threept}
\end{align}
Using \eqref{eq:disc_gen_4pt}, we find that the discontinuity of $\psi^{\f'^3}_{4}$ is,
\begin{align}
\text{disc}_{s}\psi^{\f'^3}_{4}(k_a,s,t)=8 g_1^2 (k_1 k_2 k_3 k_4)^2 s^{3/2} \frac{\left(3 (k_1+k_2)^2+s\right) \left(3 (k_3+k_4)^2+s\right)}{\left((k_1+k_2)^2-s\right)^3 \left((k_3+k_4)^2-s\right)^3}.
\end{align}
To evaluate the integral in \eqref{eq:dispphipcubed}, we follow the same procedure used to evaluate $\psi^{\f\f'^2}_{4}$: We make the change of variables $s'=p^2$, extend the integration of $p$ to $(-\infty,\infty)$, and then close the contour in either the upper or lower half-plane. This yields,
\begin{align}
\psi^{\f'^3}_{4}(k_a,s,t)&=\frac{1}{2\pi i}\int\limits_{-\infty}^{\infty}\frac{dp}{p^2-s+i\epsilon}\frac{8 g_1^2 k_1^2 k_2^2 k_3^2 k_4^2 p^4 \left(3 (k_1+k_2)^2+p^2\right) \left(3 (k_3+k_4)^2+p^2\right)}{\left((k_1+k_2)^2-p^2\right)^3 \left((k_3+k_4)^2-p^2\right)^3}
\nonumber\\
&=\frac{4 g_1^2 k_1^2 k_2^2 k_3^2 k_4^2 }{E_L^3 E_R^3 k_T^5}\bigg(s \big(6 E_L^2 E_R^2+k_T^3 (E_L+E_R)+k_T^2 (E_L+E_R)^2
\nonumber
\\
&\hspace{1.3in} +3 E_L E_R k_T (E_L+E_R)+k_T^4\big)-6 E_L^3 E_R^3\bigg),\label{eq:finalEFT1squared}
\end{align}
where $E_{L,R}$ are the partial energies in the $s$-channel,
\begin{align}
E_L=k_1+k_2+\sqrt{s}, \qquad E_R=k_3+k_4+\sqrt{s},
\end{align}
and $k_T=k_1+\ldots+k_4$ is the total energy at four points.
We can note that the poles in $p$ in the first line of \eqref{eq:finalEFT1squared} are in the same location as in \eqref{eq:disp_phiphipsq_Pt2}.
In the next section we will show that this pattern holds to all orders in the EFT of inflation.

As a consistency check, the result \eqref{eq:finalEFT1squared} passes the manifest locality test \eqref{eq:MLT_test} \cite{Jazayeri:2021fvk} and is regular in the folded limit.
Once again, it is expected that these consistency conditions hold since we have directly rewritten the Feynman diagram expression \eqref{eq:EFT1sqFeynman} using the bulk dispersion formula \eqref{eq:bulkdispersion}.

A more non-trivial example is the exchange diagram constructed from two $\f'(\partial_i\f)^2$ vertices.
In this case we will show that the dispersion formula and Feynman diagram results differ by a finite sum of quartic contact diagrams.
To start, we need the three-point wavefunction coefficient for the $\f'(\partial_i\f)^2$ interaction,
\begin{align}
\psi^{\f'(\partial_i\f)^2}_{3}(k_1,k_2,k_3)=\frac{g_2}{2e_1^3}\left(e_1^6-3 e_1^4 e_2+11 e_1^3 e_3-4 e_1^2 e_2^2-4 e_1 e_2 e_3+12 e_3^2\right).\label{eq:3ptdiscEFT2}
\end{align}
We can then use the factorization condition \eqref{eq:disc_gen_4pt} to fix the discontinuity of the exchange diagram in terms of the discontinuity of the three-point function \eqref{eq:3ptdiscEFT2},
\begin{align}
\text{disc}_{s}\psi^{\f'(\partial_i\f)^2}_{4}(k_a,s,t)=&\ P_{3/2}(\sqrt{s})\text{disc}_{s}\psi^{\f'(\partial_i\f)^2}_{3}(k_1,k_2,\sqrt{s})\text{disc}_{s}\psi^{\f'(\partial_i\f)^2}_{3}(k_3,k_4,\sqrt{s}),
\\
\text{disc}_{s}\psi^{\f'(\partial_i\f)^2}_{3}(k_1,k_2,\sqrt{s})=& \ \frac{ g_2 s^{3/2}}{\left((k_1+k_2)^2-s\right)^3}\bigg(
s^3-s^2 \left(7 k_1^2+6 k_1 k_2+7 k_2^2\right)\label{eq:discEFT2sq}
\nonumber
\\
&\hspace{.45in}
+s \left(11 k_1^4+36 k_1^3 k_2+54 k_1^2 k_2^2+36 k_1 k_2^3+11 k_2^4\right)
\nonumber
\\
&\hspace{.45in}
-(k_1+k_2)^2 \big(5 k_1^4+20 k_1^3 k_2-14 k_1^2 k_2^2+20 k_1 k_2^3+5 k_2^4\big)
\bigg).
\nonumber
\\
\end{align}
For this diagram there is a new subtlety when evaluating the dispersion integral,
\begin{align}
\psi^{\f'(\partial_i\f)^2}_{4,\text{disp}}(k_a,s,t)=\frac{1}{2\pi i}\int\limits_{0}^{\infty}\frac{ds'}{s'-s+i\epsilon}\text{disc}_{s'}\psi^{\f'(\partial_i\f)^2}_{4}(k_a,s',t).\label{eq:disp_EFT2_pt1}
\end{align}
The subtlety is that $\text{disc}_{s}\psi^{\f'(\partial_i\f)^2}_{4}$ does not decay in the large $s$ limit,
\begin{align}
\lim\limits_{s\rightarrow\infty}\text{disc}_{s}\psi^{\f'(\partial_i\f)^2}_{4}(k_a,s,t)\approx g_2^{2}\left(\frac{s^{3/2}}{2}-2s^{1/2}(k_1^2+k_2^2+k_3^2+k_4^2)\right),\label{eq:growing_terms}
\end{align}
where on the right hand side we have only kept the growing terms.
Therefore, the original integral \eqref{eq:disp_EFT2_pt1} does not converge as currently defined.
The right hand side of \eqref{eq:growing_terms} is analytic in a subset of the external momenta $\mathbf{k}_a$ and is therefore semi-local in position space.
That is, it has support when a subset of the external points are coincident.\footnote{For more discussion of semi-local terms in CFTs see \cite{Bzowski:2014qja,Dymarsky:2014zja,Meltzer:2021bmb}.}
The semi-local terms in \eqref{eq:growing_terms} appear because we set $\eta_0=0$ when computing the three-point function \eqref{eq:3ptdiscEFT2}.
One way to remove these growing terms is to instead work with three-point functions that are regulated by a non-zero, late-time cutoff, $\eta_0\neq 0$.
When we compute the discontinuity $\text{disc}_{s}\psi^{\f'(\partial_i\f)^2}_{4}$ with the regulated three-point functions, we find that the dispersion integral \eqref{eq:disp_EFT2_pt1} converges.
We discuss this computation in Appendix \ref{sec:Reg_Dispersion}.

An alternative option is to define the integral \eqref{eq:disp_EFT2_pt1} by analytic continuation. 
To do this properly, we should add and subtract a finite number of terms until the integral becomes well-defined.
In Appendix \ref{sec:Reg_Dispersion} we will carry out this procedure and show that the two ways of regularizing the dispersion integral, either by using a hard cutoff or through analytic continuation, differ by a finite sum of contact diagrams.
In practice, defining the integral by analytic continuation is equivalent to making the same change of variables as before, $s'=p^2$, closing the $p$-contour in the lower half-plane, and dropping the arc at infinity.
This gives,
\begin{align}
\psi^{\f'(\partial_i\f)^2}_{4,\text{disp}}(k_a,s,t)=&\ \frac{1}{2\pi i}\int\limits_{-\infty}^{\infty}\frac{dp}{p^2-s+i\epsilon}p\hspace{.05cm}\text{disc}_{p^2}\psi^{\f'(\partial_i\f)^2}_{4}(k_a,p^2,t)
\nonumber
\\
=& \ -\left(\res\limits_{p=k_1+k_2}+\res\limits_{p=k_3+k_4}+\res\limits_{p=\sqrt{s-i\epsilon}}\right)\frac{p}{p^2 -s+i\epsilon}\text{disc}_{p^2}\psi^{\f'(\partial_i\f)^2}_{4}(k_a,p^2,t).
\label{eq:EFTsqcontour}
\end{align}
We will give the full result for \eqref{eq:EFTsqcontour} in Appendix \ref{app:Four_Point}.

As a consistency check, we can show that $\psi^{\f'(\partial_i\f)^2}_{4,\text{disp}}$ differs from the result computed in \cite{Jazayeri:2021fvk}, which we label as $\psi^{\f'(\partial_i\f)^2}_{4,\text{MLT}}$, by a finite sum of contact diagrams. 
Their result was found by using a combination of the Cosmological Optical theorem, a partial energy recursion relation, and the manifestly local test \cite{Jazayeri:2021fvk}. 
Their result in turn differs from the Feynman diagram result $\psi^{\f'(\partial_i\f)^2}_{4,\text{Feyn}}$ by a finite sum of contact diagrams.
Therefore, the output of the dispersion formula \eqref{eq:EFTsqcontour} will also differ from the Feynman diagram result by a linear combination of contact diagrams.

The four-point wavefunction coefficient found in \cite{Jazayeri:2021fvk} is given by,
\begin{align}
\psi^{\f'(\partial_i\f)^2}_{4,\text{MLT}}=-\frac{1}{2}\psi^{\f'(\partial_i\f)^2}_{\text{Res}}+g_2^2\left(5 k_T s +\frac{12 s \left(k_1^2 k_2^2+k_3^2 k_4^2\right)}{k_T^3}- \frac{25}{4}s^{3/2}-\frac{4 s^2}{k_T} \right),
\end{align}
where $\psi^{\f'(\partial_i\f)^2}_{\text{Res}}$ can be found in Appendix C of \cite{Jazayeri:2021fvk}.\footnote{To convert notations recall that $s_{\text{here}}=s^{2}_{\text{there}}$. The overall minus sign multiplying $\psi^{\f'(\partial_i\f)^2}_{\text{Res}}$ is due to a difference in conventions for the wavefunction. We also multiplied by a factor of $1/2$ to be consistent with the Cosmological Optical theorem \cite{Goodhew:2020hob} or equivalently \eqref{eq:disc_gen_4pt}. We thank the authors of \cite{Jazayeri:2021fvk} for confirming this overall factor.}
After summing over $s$-, $t$-, and $u$-channel exchange we find,
\begin{align}
\psi^{\f'(\partial_i\f)^2}_{4,\text{disp}}-\psi^{\f'(\partial_i\f)^2}_{4,\text{MLT}}+\text{(crossed-channels)}=g_2^2\left(18\psi^{\f\f'^3}_4-2\psi^{\f^2(\partial_i\f)^2}_{4}-18\psi^{\f^2\f'^2}_4-27\psi^{\f'^4}_4\right).\label{eq:diff_quartic}
\end{align}
This proves that our result for the $\f'(\partial_i\f)^2$ exchange diagram is consistent with the result of \cite{Jazayeri:2021fvk}.
We give the explicit form of the contact diagrams that appear on the right hand side of \eqref{eq:diff_quartic} in Appendix \ref{app:Four_Point}.

Finally, we can also use the factorization condition \eqref{eq:disc_gen_4pt} and dispersion formula \eqref{eq:ambiguities} to compute the two exchange diagrams that are constructed from one $\f'^3$ vertex and one $\f'(\partial_i\f)^2$ vertex.
The discontinuity of these two diagrams is,
\begin{align}
\text{disc}_{s}\psi^{\text{cross}}_{4}(k_a,s,t)=&\  P_{3/2}(\sqrt{s})\text{disc}_{s}\psi^{\f'^3}_{3}(k_1,k_2,\sqrt{s})\text{disc}_{s}\psi^{\f'(\partial_i\f)^2}_{3}(k_3,k_4,\sqrt{s})
\nonumber
\\
&+(k_1,k_2)\leftrightarrow (k_3,k_4).
\end{align}
The computation of the full wavefunction coefficient using the dispersion formula is unchanged in comparison to the previous examples.
The explicit form of $\psi^{\text{cross}}_{4}$ is given in Appendix \ref{app:Four_Point}.

\subsubsection{All Orders}
We will now explain how to generalize the previous results to all orders in the EFT of inflation.
Our starting point is the following ansatz for the three-point wavefunction coefficient of a single, massless, self-interacting scalar $\phi$ \cite{Jazayeri:2021fvk}:
\begin{align}
\psi^{(r)}_{3}(k_1,k_2,k_3)=\frac{1}{e_1^r}\sum\limits_{n=0}^{\lfloor\frac{r+3}{3}\rfloor}\sum\limits_{m=0}^{\lfloor\frac{r+3-3n}{2}\rfloor}C_{mn}e_{1}^{3+r-2m-3n}e_2^me_3^n.\label{eq:threeptansatz}
\end{align}
The form of this ansatz is fixed by using a set of boostless bootstrap rules \cite{Pajer:2020wxk}.
The label $r$ in $\psi^{(r)}_{3}$ corresponds to the strength of the total energy pole. 
Here we recall that $e_1$ is the total energy variable at three points, $e_1=k_1+k_2+k_3$.
The previous examples considered in this section correspond to different linear combinations of $\eqref{eq:threeptansatz}$ for $r\leq 3$.
For $r=0$ the ansatz \eqref{eq:threeptansatz} needs to be modified to include a logarithmic term that appears in $\f^3$ theory.\footnote{A local non-Gaussianity gives a rational three-point function with $r=0$ \cite{Jazayeri:2021fvk}.}
Here we are interested in interactions that break dS boosts, so we will ignore any logarithmic modification of the ansatz \eqref{eq:threeptansatz}.

In \cite{Jazayeri:2021fvk} they used the manifestly local test to derive recursion relations for the coefficients $C_{mn}$.
This test can be used to find all three-point wavefunction coefficients that come from a manifestly local Lagrangian in dS. 
In this section we will not need the explicit form of the coefficients $C_{mn}$, but we will use the manifestly local test to classify what poles contribute to the dispersion integral.

The discontinuity of a general exchange diagram in the EFT of inflation takes the form,
\begin{align}
\disc_{s}\psi^{(r_1,r_2)}(k_a,s,t)=P_{3/2}(\sqrt{s})\text{disc}_{s}\psi^{(r_1)}_3(k_1,k_2,\sqrt{s})\text{disc}_{s}\psi^{(r_2)}_3(k_3,k_4,\sqrt{s}).\label{eq:discEFT_inflation}
\end{align}
From \eqref{eq:threeptansatz} we see that the right hand side of \eqref{eq:discEFT_inflation} has poles at,
\begin{align}
s=(k_1+k_2)^2,\label{eq:pole_generic_1}
\\
s=(k_3+k_4)^2.\label{eq:pole_generic_2}
\end{align}

One may also expect a singularity at $s=0$ since $P_{3/2}(\sqrt{s})\propto s^{-3/2}$, but it was shown in \cite{Jazayeri:2021fvk} that the right hand side of \eqref{eq:discEFT_inflation} vanishes as $s\rightarrow 0$.
To show this, one uses that the ansatz \eqref{eq:threeptansatz} for $\psi^{(r)}_3(k_1,k_2,k_3)$ has a Taylor series expansion around $k_3=0$. The manifestly local test implies that the linear term in this expansion vanishes,
\begin{align}
\partial_{k_3}\psi^{(r)}_3(k_1,k_2,k_3)\bigg|_{k_3=0}=0.
\end{align}
In addition, taking the discontinuity of $\psi^{(r)}_3(k_1,k_2,k_3)$ in $k_3^2$ projects out terms that are even in $k_3$.
These two conditions imply the following small-$s$ scaling, 
\begin{align}
\lim\limits_{s\rightarrow 0}\text{disc}_{s}\psi^{(r)}_3(k_1,k_2,\sqrt{s})\sim s^{3/2}.
\end{align}
This shows that the right hand side of \eqref{eq:discEFT_inflation} scales like $s^{3/2}$ as $s\rightarrow 0$.
Therefore, \eqref{eq:discEFT_inflation}  only has poles at the locations \eqref{eq:pole_generic_1} and \eqref{eq:pole_generic_2}.

Next we study the dispersion integral,
\begin{align}
\psi^{(r_1,r_2)}_{\text{disp}}(k_a,s,t)=\frac{1}{2\pi i}\int\limits_{0}^{\infty}\frac{ds'}{s' - s+i\epsilon}\disc_{s'}\psi^{(r_1,r_2)}(k_a,s',t).\label{eq:gen_disp}
\end{align}
From the ansatz \eqref{eq:threeptansatz} and the general form of the discontinuity \eqref{eq:discEFT_inflation}, we see that the integrand only contains half-integer powers of $s$.
After making the standard change of variables, $s'=p^2$, we find that the integrand is an even function of $p$ and can be rewritten as,
\begin{align}
\psi^{(r_1,r_2)}_{\text{disp}}(k_a,s,t)=\frac{1}{2\pi i}\int\limits_{-\infty}^{\infty}\frac{dp}{p^2 -s+i\epsilon}p\hspace{.05cm} \disc_{p^2}\psi^{(r_1,r_2)}(k_a,p^2,t).
\end{align}
Finally, we can close the $p$-contour in the lower half-plane and pick up the poles at,
\begin{align}
p&=k_1+k_2,
\\
p&=k_3+k_4,
\\
p&=\sqrt{s-i\epsilon},
\end{align}
where we recall that the $k_a$ implicitly have a small, negative imaginary part.
This yields,
\begin{align}
\psi^{(r_1,r_2)}_{\text{disp}}(k_a,s,t)&=-\left(\res\limits_{p=k_1+k_2}+\res\limits_{p=k_3+k_4}+\res\limits_{p=\sqrt{s-i\epsilon}}\right)\frac{p}{p^2 -s+i\epsilon}\disc_{p^2}\psi^{(r_1,r_2)}(k_a,p^2,t)
\nonumber\\[5pt]
&=-\left(\res\limits_{p=k_1+k_2}+\res\limits_{p=k_3+k_4}+\res\limits_{p=\sqrt{s-i\epsilon}}\right)\frac{p}{p^2 -s+i\epsilon}P_{3/2}(p)\text{disc}_{p^2}\psi^{(r_1)}_3(k_1,k_2,p)
\nonumber \\
&\hspace{3.39in}\times\text{disc}_{p^2}\psi^{(r_2)}_3(k_3,k_4,p).\label{eq:4pt_arbr1r2}
\end{align}
This expression holds for arbitrary $r_1$ and $r_2$.
If the original dispersion integral \eqref{eq:gen_disp} diverges and needs to be regularized, then the expression \eqref{eq:4pt_arbr1r2} corresponds to defining the integral by analytic continuation.
More details on this procedure can be found in Appendix \ref{sec:Reg_Dispersion}.

As an example, we can consider a diagram involving two $\f\f''^2$ interactions.
The three-point wavefunction coefficient and its discontinuity are given by,
\begin{align}
\psi_{3}^{\f\f''^2}(k_1,k_2,k_3)=&\frac{e_2^3 e_1+3 e_2^2 e_3-3 e_2 e_3 e_1^2}{e_1^4},
\\
\disc_{s}\psi_{3}^{\f\f''^2}(k_1,k_2,\sqrt{s})=&\frac{2 s^{3/2}}{\left((k_1+k_2)^2-s\right)^4}\bigg(3 s^2 \left(k_1^2 k_2^2-k_1^4-k_2^4\right)+\big(k_1^6+6 k_1^5 k_2+10 k_1^3 k_2^3
\nonumber
\\
&
+6 k_1 k_2^5+k_2^6\big) (k_1+k_2)^2+2 s \big(k_1^6+12 k_1^5 k_2+24 k_1^4 k_2^2+28 k_1^3 k_2^3
\nonumber
\\
&
+24 k_1^2 k_2^4+12 k_1 k_2^5+k_2^6\big)\bigg).
\end{align}
The four-point wavefunction coefficient is then,
\begin{align}
\psi^{\f\f''^2}_{4,\text{disp}}(k_a,s,t)&=-\left(\res\limits_{p=k_1+k_2}+\res\limits_{p=k_3+k_4}+\res\limits_{p=\sqrt{s-i\epsilon}}\right)\frac{p}{p^2 -s+i\epsilon}P_{3/2}(p)\text{disc}_{p^2}\psi^{\f\f''^2}_3(k_1,k_2,p)
\nonumber \\
&\hspace{3.4in}\times\text{disc}_{p^2}\psi^{\f\f''^2}_3(k_3,k_4,p).
\end{align}
It is now straightforward to compute the $\f\f''^2$ exchange diagram using the above three-point function.
In this example, the dispersion integral \eqref{eq:gen_disp} converges and does not require an additional regularization.

\subsection{Five-Point Function}
As a final example, we will consider a five-point diagram constructed from three $\f'^3$ vertices.
This was recently computed in \cite{Goodhew:2021oqg} and we will show how to recover their result using the dispersion formula.

The full Feynman diagram is,
\begin{align}
\psi^{\f'^3}_{5}(k_a,s_{12},s_{45})=& \hspace{.1cm} \begin{aligned}
		\includegraphics[scale=.3]{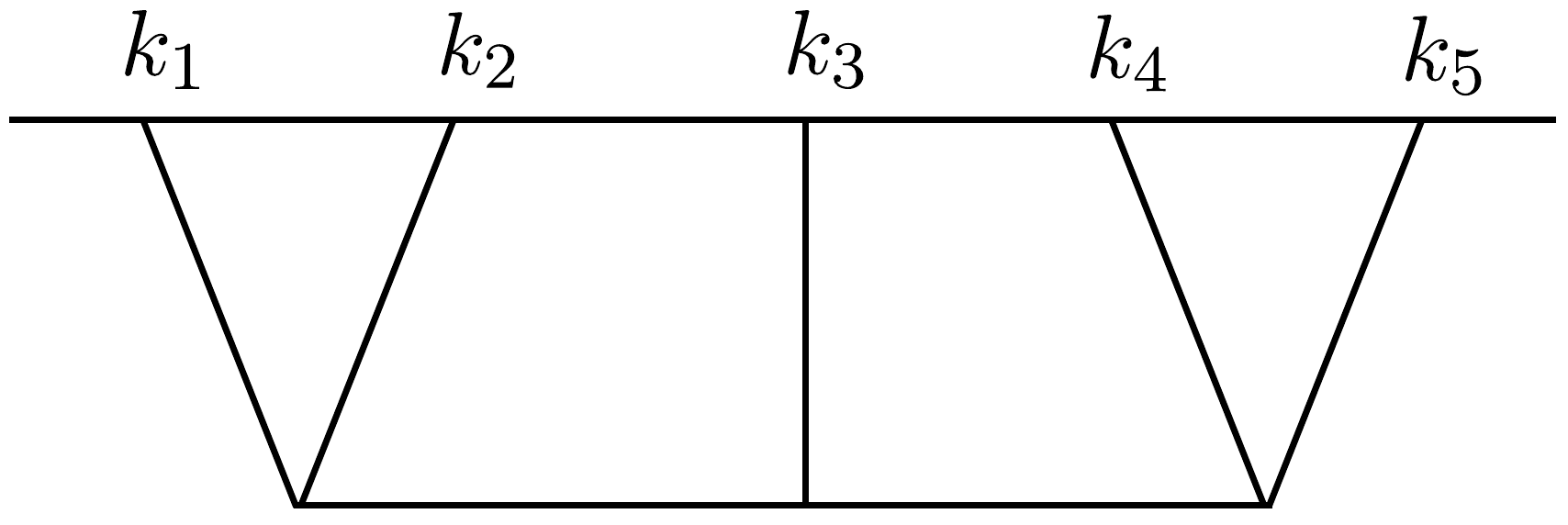}
	\end{aligned}
\nonumber
\\
=& \ 
-(ig_1)^3\int\limits_{-\infty}^{0}\frac{d\eta_1d\eta_2d\eta_3}{\eta_1\eta_2\eta_3}\mathcal{K}'_\f(k_1,\eta_1)\mathcal{K}'_\f(k_2,\eta_1)\partial_{\eta_1}\partial_{\eta_2}\mathcal{G}_\f(\sqrt{s_{12}},\eta_1,\eta_2)
\nonumber
\\
&\hspace{1.56in}\mathcal{K}'_\f(k_3,\eta_2)\partial_{\eta_2}\partial_{\eta_3}\mathcal{G}_\f(\sqrt{s_{45}},\eta_2,\eta_3)\mathcal{K}'_\f(k_4,\eta_3)\mathcal{K}'_\f(k_5,\eta_3),\label{eq:five_point_explicit}
\end{align}
where we use a prime to denote $\partial_{\eta}$ where it is unambiguous. 
The only $s_{12}$ dependence comes from the propagator and therefore it is trivial to write down a dispersion formula for \eqref{eq:five_point_explicit} using the bulk dispersion formula \eqref{eq:bulkdispersion}:
\begin{align}
\psi^{\f'^3}_{5}(k_a,s_{12},s_{45})=\frac{1}{2\pi i}\int\limits_{0}^{\infty}\frac{ds'_{12}}{s'_{12}-s_{12}+i\epsilon}\text{disc}_{s_{12}'}\psi^{\f'^3}_{5}(k_a,s_{12}',s_{45}).\label{eq:disp_5pt_pt1}
\end{align}
Taking the discontinuity in $s_{12}$ at fixed $k_a$ and $s_{45}$ factorizes the five-point diagram into a product of three- and four-point diagrams,
\begin{align}
\text{disc}_{s_{12}}\psi^{\f'^3}_{5}(k_a,s_{12},s_{45})&= \ P_{3/2}(\sqrt{s_{12}})
\begin{aligned}
		\includegraphics[scale=.3]{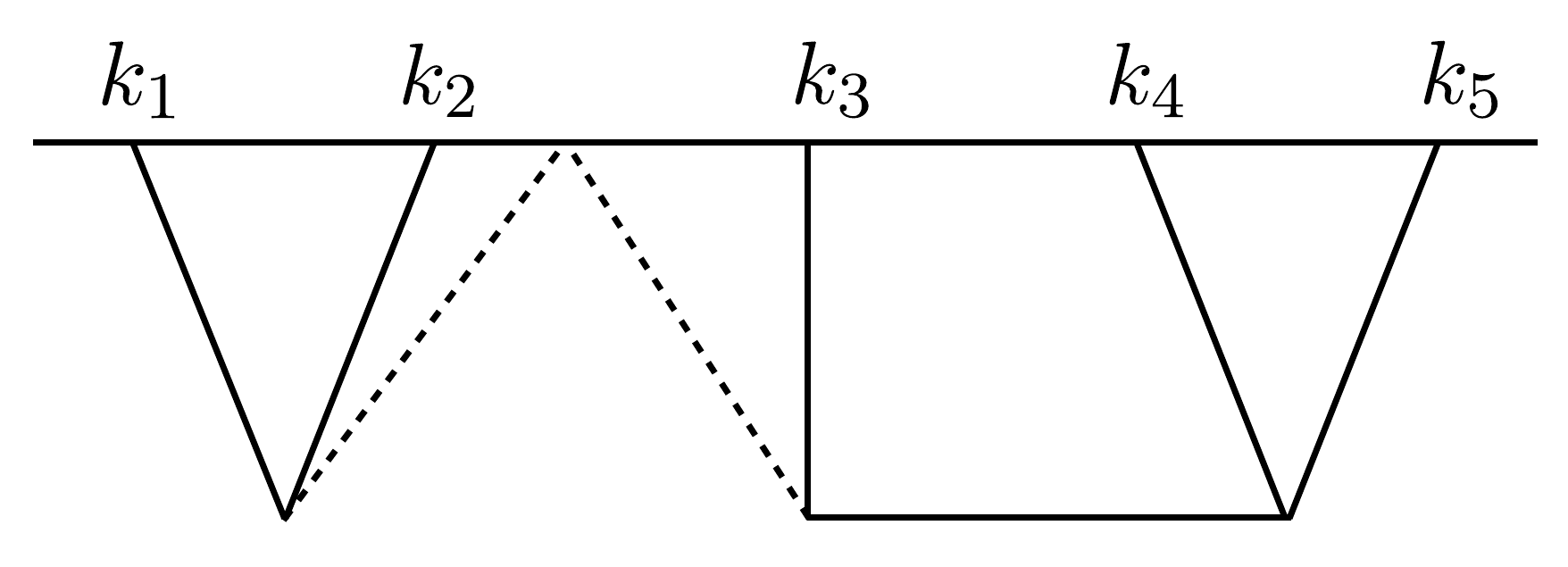}
\end{aligned}
\nonumber
\\
&=P_{3/2}(\sqrt{s_{12}})\disc_{s_{12}}\psi^{\f'^3}_{3}(k_1,k_2,\sqrt{s_{12}})
\disc_{s_{12}}\psi^{\f'^3}_{4}(\sqrt{s_{12}},k_3,k_4,k_5,\sqrt{s_{45}}). \label{eq:disc5pt}
\end{align}
The discontinuity \eqref{eq:disc5pt} can be computed using the known results for the three-point function, \eqref{eq:EFT1threept}, and the four-point exchange diagram, \eqref{eq:finalEFT1squared}.
Using these results, we find:
\begin{align}
\disc_{s_{12}}\psi^{\f'^3}_{5}=\ & 4 g_1^3 (k_1 k_2 k_3 k_4 k_5)^2 q_1 \left(\frac{1}{(k_1+k_2+q_1)^3}-\frac{1}{(k_1+k_2-q_1)^3}\right)
\nonumber
\\
&\hspace{-.5cm}
\bigg\{\frac{6}{(k_3+k_4+k_5+q_1)^5}-\frac{6}{(k_3+k_4+k_5-q_1)^5}+ \frac{1}{(k_4+k_5-q_2)^3 (k_4+k_5+q_2)^3}
\nonumber
\\
&\hspace{-.5cm}
\bigg[
q_2^2 \bigg(\frac{3 (k_4+k_5)^2+q_2^2}{(k_3+k_4+k_5+q_1)^3}+\frac{3 (k_4+k_5)^2+q_2^2}{(k_3-q_1+q_2)^3}-\frac{3 (k_4+k_5)^2+q_2^2}{(k_3+k_4+k_5-q_1)^3}
\nonumber
\\
&\hspace{-.5cm}
-\frac{3 (k_4+k_5)^2+q_2^2}{(k_3+q_1+q_2)^3}+\frac{6 (k_4+k_5-q_2)^2 (k_4+k_5+q_2)^2}{(k_3+k_4+k_5+q_1)^5}-\frac{6 (k_4+k_5-q_2)^2 (k_4+k_5+q_2)^2}{(k_3+k_4+k_5-q_1)^5}
\nonumber
\\
&\hspace{-.5cm}
+\frac{6 (k_4+k_5) (k_4+k_5-q_2) (k_4+k_5+q_2)}{(k_3+k_4+k_5+q_1)^4}-\frac{6 (k_4+k_5) (k_4+k_5-q_2) (k_4+k_5+q_2)}{(k_3+k_4+k_5-q_1)^4}\bigg)\bigg]
\bigg\},\label{eq:disc5ptV2}
\end{align}
where we defined $q_1=\sqrt{s_{12}}$ and $q_2=\sqrt{s_{45}}$ to make the expression more compact.
The result \eqref{eq:disc5ptV2} agrees with what was found in \cite{Goodhew:2021oqg} using a different discontinuity.

To evaluate the dispersion integral \eqref{eq:disp_5pt_pt1}, we perform the change of variables $s'_{12}=p^2$ and rewrite the dispersion integral as a contour integral:
\begin{align}
\psi^{\f'^3}_{5}(k_a,s_{12},s_{45})=\frac{1}{2\pi i}\int\limits_{-\infty}^{\infty}\frac{dp}{p^2-s_{12}+i\epsilon}p\hspace{.05cm} \text{disc}_{p^2}\psi^{\f'^3}_{5}(k_a,p^2,s_{45}).\label{eq:five_pt_int_Pt2}
\end{align}
The integrand of \eqref{eq:five_pt_int_Pt2} has poles in $p$ at:
\begin{align}
p&=\pm(k_1+k_2),
\\
p&=\pm(k_3+k_4+k_5),
\\
p&=\pm(k_3+\sqrt{s_{45}}),
\\
p&=\pm\sqrt{s_{12}-i\epsilon}.
\end{align}
Closing the contour in the lower half-plane yields,
\begin{align}
\psi^{\f'^3}_{5}=-\bigg(&\res\limits_{p=k_1+k_2}+\res\limits_{p=k_3+k_4+k_5}
\nonumber \\
&\hspace{.5in} +\res\limits_{p=k_3+\sqrt{s_{45}}}+\res\limits_{p=\sqrt{s_{12}-i\epsilon}}\bigg)\frac{p}{p^2-s_{12}+i\epsilon}\text{disc}_{p^2}\psi^{\f'^3}_{5}.\label{eq:5pt_sum_res}
\end{align} 
We can then compute the five-point diagram using \eqref{eq:disc5ptV2} and \eqref{eq:5pt_sum_res}, although we will not write out the full expression here.

\section{An AdS Perspective}
\label{sec:AdSPerspective}
In this section we will explain how the dS unitarity methods studied in the previous sections are related by analytic continuation to AdS unitarity methods.

\subsection{AdS/CFT Perturbation Theory}
We will study a single scalar field $\Phi(\mathbf{x},z)$ in the Poincar\'e patch of AdS$_{d+1}$,
\begin{align}
ds^{2}_{\text{AdS}}=\frac{1}{z^2}(dz^2+\eta_{\mu\nu}dx^{\mu}dx^{\nu}),
\end{align}
where $\eta_{\mu\nu}$ is the $d$-dimensional Minkowski metric and we set the AdS radius to one, $R_{\text{AdS}}=1$.
The radial direction $z$ lies in the range $z\in(z_{c},\infty)$, where $z_c$ is the bulk IR cutoff.
The mass $m$ of the bulk field $\Phi$ is related to the dimension $\Delta$ of the boundary CFT operator $\f$ by:
\begin{align}
\Delta(\Delta-d)=m^2.
\end{align}
In this section we will assume that $\Delta$ is real.
As in dS, we will parameterize the conformal dimensions using $\Delta=d/2+\nu$.

We will study time-ordered correlation functions in momentum space, $\<T[\f(\mathbf{k}_1)\ldots\f(\mathbf{k}_n)]\>$. Here $T$ is the time-ordered product.
It is convenient to factor out the overall momentum conserving $\delta$-function using a double-bracket notation,
\begin{align}
\<T[\f(\mathbf{k}_1)\ldots \f(\mathbf{k}_n)]\>=(2\pi)^d\delta(\mathbf{k}_1+\ldots+\mathbf{k}_n) \llc T[\f(\mathbf{k}_1)\ldots \f(\mathbf{k}_n)]\rrc.
\end{align}
The bulk-to-boundary and bulk-to-bulk propagators for a scalar in the Poincar\'e patch are,
\begin{align}
\mathcal{K}^{\text{AdS}}_{\nu}(k,z,z_c)&=\left(\frac{z}{z_c}\right)^{\frac{d}{2}}\frac{K_{\nu}(kz)}{K_{\nu}(kz_c)}, \label{eq:GBbAdS}
\\
\mathcal{G}^{\text{AdS}}_{\nu}(k,z_1,z_2,z_c)=&-i(z_1z_2)^{\frac{d}{2}}\bigg(\theta(z_1-z_2)K_{\nu}(kz_1)I_{\nu}(kz_2)+\theta(z_2-z_1)K_{\nu}(kz_2)I_{\nu}(kz_1)
\nonumber 
\\ &\hspace{.9in}-K_{\nu}(kz_1)K_{\nu}(kz_2)\frac{I_{\nu}(kz_c)}{K_{\nu}(kz_c)}\bigg),\label{eq:GBBAdS}
\end{align}
where $I_\nu$ and $K_\nu$ are the modified Bessel functions of the first and second kind, respectively.
The last term in \eqref{eq:GBBAdS} comes from imposing the Dirichlet boundary conditions at $z=z_c$.
In \eqref{eq:GBbAdS} and \eqref{eq:GBBAdS} we have set the speed of propagation in the radial direction to one, $c_r=1$.
This is the analog of setting $c_s=1$ in the dS action \eqref{eq:freeactiondS}.
As in Section \ref{sec:dSFeynman}, we can always restore the dependence on the radial speed $c_r$ by using dimensional analysis.

The AdS propagators used to compute time-ordered correlators are related by analytic continuation in $k$ to the dS propagators used to compute the wavefunction \cite{Maldacena:2002vr,Harlow:2011ke,Isono:2020qew,McFadden:2009fg,McFadden:2010na},\footnote{Note that here we are relabelling the variable, $\eta\rightarrow-z$, and are not analytically continuing in $\eta$.}
\begin{align}
\mathcal{K}_{\nu}^{\text{AdS}}(k,z,z_c)\bigg|_{d=3}&=\mathcal{K}_{\nu}^{\text{dS}}(e^{-\frac{i\pi}{2}}k,-z,-z_c),\label{eq:AdStodSPt1}
\\[5pt]
\mathcal{G}_{\nu}^{\text{AdS}}(k,z_1,z_2,z_c)\bigg|_{d=3}&=-\mathcal{G}_{\nu}^{\text{dS}}(e^{-\frac{i\pi}{2}}k,-z_1,-z_2,-z_c),\label{eq:AdStodSPt2}
\end{align} 
where $-\pi/2<\text{arg}\ k\leq\pi$.
In the remainder of this section we will drop the explicit dependence on the cutoffs $z_c$ and $\eta_0$.

It is also possible to continue the analogy with Section \ref{sec:dSFeynman} by writing down interactions which preserve scale and Poincar\'e invariance in the dual QFT, but break special conformal transformations.
These types of interactions appear in the EFT of slow holographic RG flows \cite{Kaplan:2014dia,Baumann:2019ghk}.\footnote{As with inflation, in the EFT of slow holographic RG flows scale invariance is only an approximate symmetry. It is broken by the analog of slow-roll effects \cite{Kaplan:2014dia,Baumann:2019ghk}. For a review on scale vs conformal invariance see \cite{Nakayama:2013is}.}
They have the same schematic form as \eqref{eq:gen_boost_breaking_action},
\begin{align}
S_{\text{int}}=\int dz d^{d}\mathbf{x}\ z^{\sum_aN_a-d-1}\partial^{N_1}\Phi\ldots\partial^{N_n}\Phi.\label{eq:AdSinteractions}
\end{align}
The only difference is that here $\partial$ denotes a derivative with respect to $z$ or $x^{\mu}$. 
As before in dS, $N_a$ counts the total number of derivatives acting on the fields and the power of $z$ is fixed by imposing scale invariance.
There is then a one-to-one correspondence between interactions that appear in the EFT of inflation and the EFT of slow holographic RG flows \cite{Kaplan:2014dia,Baumann:2019ghk}.

\subsection{AdS Dispersion Formulas}
In this section we will prove the analogs of the tree-level factorization condition \eqref{eq:disc_gen_4pt} and dispersion formula \eqref{eq:ambiguities} in AdS for generic boost-breaking interactions.
The proof of the tree-level cutting rules in AdS will be identical in form to the dS proof given in Section \ref{sec:dSWavefunction} and the previous AdS arguments of \cite{Meltzer:2020qbr,Meltzer:2021bmb}, so we will keep this discussion brief.
We will then prove the tree-level dispersion formula in pure AdS by using Bessel function identities.
We will give a more general proof of the dispersion formula in asymptotically AdS spacetimes in Appendix \ref{app:dispersion}.

To derive the analog of \eqref{eq:disc_gen_4pt} in AdS, we study the discontinuity of the AdS bulk-to-bulk propagator.
Using either the explicit form of the bulk-to-bulk propagator \eqref{eq:GBBAdS} or its relation to the dS propagator \eqref{eq:AdStodSPt2}, we can show:
\begin{align}
\text{disc}_{s_-} \mathcal{G}_{\nu}^{\text{AdS}}(k,z_1,z_2)=P^{\text{AdS}}_{\nu}(k)\disc_{s_-}\mathcal{K}_{\nu}^{\text{AdS}}(k,z_1)\disc_{s_-}\mathcal{K}_{\nu}^{\text{AdS}}(k,z_2),\label{eq:AdSdiscFact}
\end{align}
where as a reminder $s_-=-\mathbf{k}^2$.\footnote{To keep the notation compact, we will use $k$ and $\sqrt{-s_-}$ interchangeably where it is unambiguous.}
The discontinuity in $s_-$ is defined by,
\begin{align}
\text{disc}_{s_-}f(s_-)\equiv f(s_-+i\epsilon)-f(s_- - i\epsilon).
\end{align}
Finally, the AdS ``power spectrum" is defined by,
\begin{align}
P^{\text{AdS}}_{\nu}(k)=\frac{\pi}{4}z_{c}^{d}H^{(1)}_{\nu}\left(e^{\frac{i\pi}{2}}kz_{c}\right)H^{(2)}_{\nu}\left(e^{\frac{i\pi}{2}}kz_{c}\right).\label{eq:AdSPower}
\end{align}
The bulk AdS identity \eqref{eq:AdSdiscFact} is related by analytic continuation to the bulk dS identity \eqref{eq:discGgen}.
As in dS, we can use \eqref{eq:AdSdiscFact} to show that the discontinuity of an AdS exchange diagram is proportional to a product of three-point functions,
\begin{align}\disc_{s_-} \
\begin{aligned}
		\includegraphics[scale=.25]{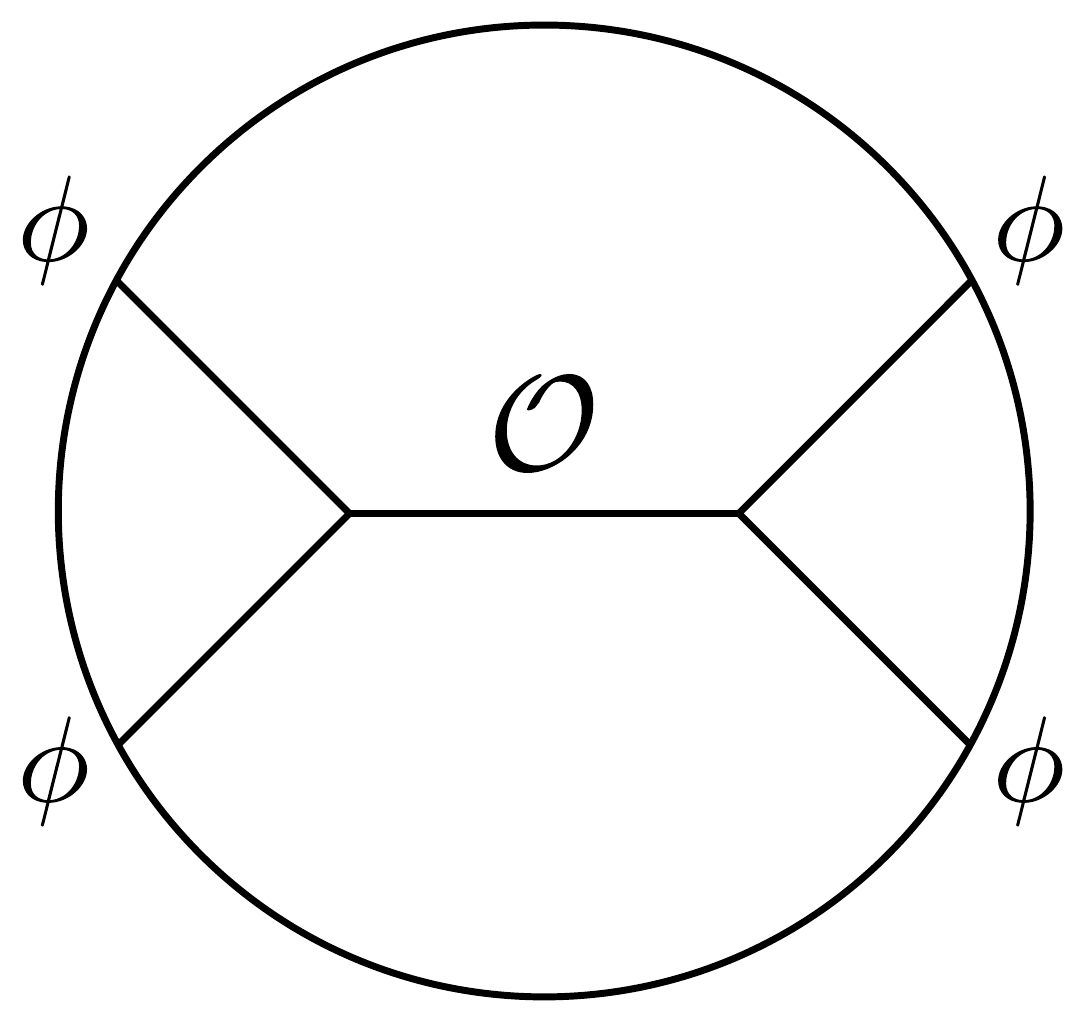}
\end{aligned}
	= P_{\nu}^{\text{AdS}}(k) \ \begin{aligned}
		\includegraphics[scale=.25]{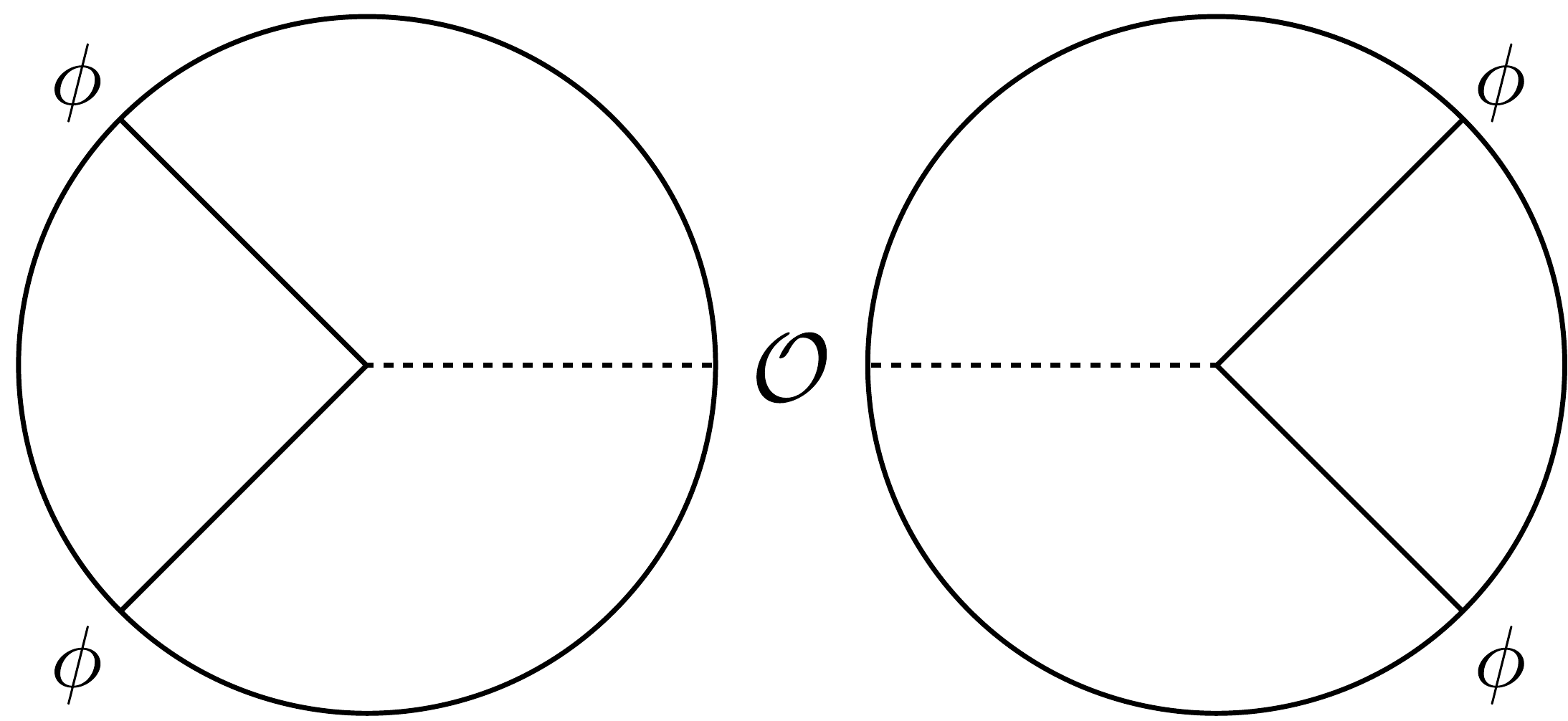}
\end{aligned}\ ,\label{eq:disc_witten_figure}
\end{align}
where $\O$ is the CFT primary operator dual to the bulk exchanged field and we use dashed lines to denote the discontinuity of the appropriate propagator.
Analytic continuation maps the AdS factorization condition \eqref{eq:disc_witten_figure} to the corresponding dS factorization condition \eqref{eq:disc_gen_4pt}.

The benefit of working in AdS is that we can explain why the discontinuity of the AdS bulk-to-bulk propagator factorizes in $(z_1,z_2)$ using general operator identities.
To show this, we recall that the AdS bulk-to-bulk propagator is defined in terms of a time-ordered, free-field two-point function:
\begin{align}
\mathcal{G}_{\nu}^{\text{AdS}}(\mathbf{x}_{12},z_1,z_2)&=\< T[\Phi(\mathbf{x}_1,z_1)\Phi(\mathbf{x}_2,z_2)]\>_{\text{free}},\label{eq:Def_G_Tord}
\end{align}
where $\mathbf{x}_{ij}=\mathbf{x}_i-\mathbf{x}_j$.
In local, causal, unitary theories, the discontinuity of the non-perturbative, time-ordered two-point function is equal to a sum of Wightman two-point functions,
\begin{align}
\disc_{s_-}\llc T[\Phi(\mathbf{k},z_1)\Phi(-\mathbf{k},z_2)]\rrc=\llc \Phi(\mathbf{k},z_1)\Phi(-\mathbf{k},z_2)\rrc +(\mathbf{k},z_1)\leftrightarrow (-\mathbf{k},z_2).\label{eq:disc_G_to_W}
\end{align}
The operators on the right hand side of \eqref{eq:disc_G_to_W} are ordered as written.
We can always factorize the Wightman two-point functions $\<\Phi_1\Phi_2\>$ by inserting a complete set of states between the two operators.
In free-field theory the identity \eqref{eq:disc_G_to_W} reduces to the previous expression \eqref{eq:AdSdiscFact}.\footnote{For a review of how to derive \eqref{eq:disc_G_to_W} in flat-space, see \cite{Meltzer:2021bmb} and references therein.
The corresponding proof in the AdS Poincar\'e patch is identical, except that we need to keep track of an additional radial variable.}

We will now prove the analog of the tree-level dispersion formula \eqref{eq:ambiguities} for boost-breaking AdS theories.
We will study the following general, scalar exchange diagram for $\<\f\f\f\f\>$,
\begin{align}
W^{\text{Feyn}}_{\O}(k_a,s_-,t_-)&=-\int\limits\frac{dz_1dz_2}{z_1^{d+1}z_2^{d+1}}V_{L}V_{R}\mathcal{K}^{\text{AdS}}_{\f}(k_1,z_1)\mathcal{K}^{\text{AdS}}_{\f}(k_2,z_1)\mathcal{G}^{\text{AdS}}_{\O}(\sqrt{-s_-},z_1,z_2)
\nonumber
\\
&\hspace{1.186in}\mathcal{K}^{\text{AdS}}_{\f}(k_3,z_2)\mathcal{K}^{\text{AdS}}_{\f}(k_4,z_2),\label{eq:AdS_exchange_diagram}
\end{align}
where $V_{L,R}$ are the left and right vertex factors and $\O$ is a generic, scalar primary.
This is the same Witten diagram drawn on the left hand side of \eqref{eq:disc_witten_figure}.
The dispersive representation of this exchange diagram is,
\begin{align}
W^{\text{disp}}_{\O}(k_a,s_-,t_-)&=\frac{1}{2\pi i}\int\limits_0^{\infty}\frac{ds_-'}{s'_--s_--i\epsilon}\disc_{s'_-}W^{\text{Feyn}}_{\O}(k_a,s'_-,t_-),
\\
\disc_{s_-}W^{\text{disp}}_{\O}(k_a,s_-,t_-)&=-\int\limits\frac{dz_1dz_2}{z_1^{d+1}z_2^{d+1}}V_{L}V_{R}\mathcal{K}^{\text{AdS}}_{\f}(k_1,z_1)\mathcal{K}^{\text{AdS}}_{\f}(k_2,z_1)
\nonumber
\\
&\hspace{1in}\disc_{s_-}\mathcal{G}^{\text{AdS}}_{\O}(\sqrt{-s_-},z_1,z_2)\mathcal{K}^{\text{AdS}}_{\f}(k_3,z_2)\mathcal{K}^{\text{AdS}}_{\f}(k_4,z_2).
\end{align}

We want to show that the difference between $W^{\text{Feyn}}_{\O}$ and $W^{\text{disp}}_{\O}$ is a sum of contact Witten diagrams.
To prove this, we will use that the vertex factors $V_{L,R}$ are polynomials in $\mathbf{k}_a\cdot \mathbf{k}_b$ and differential operators in $z_{1,2}$.
The exchange Witten diagram \eqref{eq:AdS_exchange_diagram} then has the schematic form,
\begin{align}
W^{\text{Feyn}}_{\O}(k_a,s_-,t_-)=-\int\limits\frac{dz_1dz_2}{z_1^{d+1}z_2^{d+1}}&\mathcal{D}_{L,z_1}\mathcal{D}_{R,z_2}\left(\mathcal{K}^{\text{AdS}}_{\f}(k_1,z_1)\mathcal{K}^{\text{AdS}}_{\f}(k_2,z_1)\mathcal{K}^{\text{AdS}}_{\f}(k_3,z_2)\mathcal{K}^{\text{AdS}}_{\f}(k_4,z_2)\right)
\nonumber
\\
&\text{Poly}(k_a,s_-)\mathcal{G}^{\text{AdS}}_{\O}(\sqrt{-s_-},z_1,z_2).\label{eq:AdS_exchange_diagramV2}
\end{align}
Here $\mathcal{D}_{L,R}$ are differential operators in $z$ associated to the vertex factors, $V_{L,R}$. 
We have integrated by parts so the $z$-derivatives only act on the bulk-to-boundary propagators.
The term $\text{Poly}(k_a,s_-)$ is a general polynomial in the external Lorentz invariants, $\{k_a,s_-\}$.\footnote{The polynomials are independent of $t_-$ because we are studying scalar exchange diagrams.}
We can assume without loss of generality that $\text{Poly}(k_a,s_-)$ is a monomial, $s_-^m\prod_ak_a^{n_a} $.
In that case, the difference between $W^{\text{Feyn}}_{\O}$ and $W^{\text{disp}}_{\O}$ is,
\begin{align}
W^{\text{Feyn}}_{\O}&(k_a,s_-,t_-)-W^{\text{disp}}_{\O}(k_a,s_-,t_-)
\nonumber
\\[3pt]
=&-\prod_ak_a^{n_a} \int\limits\frac{dz_1dz_2}{z_1^{d+1}z_2^{d+1}}
\mathcal{D}_{L,z_1}\mathcal{D}_{R,z_2}\left(\mathcal{K}^{\text{AdS}}_{\f}(k_1,z_1)\mathcal{K}^{\text{AdS}}_{\f}(k_2,z_1)\mathcal{K}^{\text{AdS}}_{\f}(k_3,z_2)\mathcal{K}^{\text{AdS}}_{\f}(k_4,z_2)\right)
\nonumber
\\
&\hspace{1.33in}\left(A^{\text{Feyn}}_\O(s_-,z_1,z_2)-A^{\text{disp}}_\O(s_-,z_1,z_2)\right),\label{eq:diff_Witten_Feyn_disp}
\end{align}
where we have defined the amputated diagrams,
\begin{align}
A^{\text{Feyn}}_\O(s_-,z_1,z_2)&=s_-^m \hspace{.05cm}\mathcal{G}^{\text{AdS}}_{\O}(\sqrt{s_-},z_1,z_2),
\\
A^{\text{disp}}_\O(s_-,z_1,z_2)&=\frac{1}{2\pi i}\int\limits_{0}^{\infty}\frac{dp}{p^2-s_- -i\epsilon}p^{2m+1} \hspace{.05cm} \disc_{p^2} \mathcal{G}^{\text{AdS}}_{\O}(p,z_1,z_2).\label{eq:disp_amputated_pt1}
\end{align}
In \eqref{eq:disp_amputated_pt1} we have made the change of variables $s'_-=p^2$.

To show that $W^{\text{Feyn}}_{\O}-W^{\text{disp}}_{\O}$ is equal to a sum of contact Witten diagrams, it is sufficient to show:
\begin{align}
\disc_{s_-}\left(A^{\text{Feyn}}_\O(s_-,z_1,z_2)-A^{\text{disp}}_\O(s_-,z_1,z_2)\right)&=0,\label{eq:vanishing_disc}
\\
A^{\text{Feyn}}_\O(s_-,z_1,z_2)-A^{\text{disp}}_\O(s_-,z_1,z_2)&=0, \quad \text{ if } \ z_1\neq z_2.\label{eq:vanishing_z12}
\end{align}
The first condition, \eqref{eq:vanishing_disc}, says that in position space the difference $A^{\text{Feyn}}_\O-A_\O^{\text{disp}}$ is localized at $\mathbf{x}_1=\mathbf{x}_2$. 
That is, the difference between the amputated diagrams is proportional to $\delta(\mathbf{x}_{12})$ and its derivatives.
The second condition, \eqref{eq:vanishing_z12}, similarly says that the difference between the two amputated diagrams is localized at $z_1=z_2$.
Therefore, if the above two conditions hold, then $A^{\text{Feyn}}_\O-A^{\text{disp}}_\O$ only has support at a single bulk point in position space, $(\mathbf{x}_1,z_1)=(\mathbf{x}_2,z_2)$, and \eqref{eq:diff_Witten_Feyn_disp} is equal to a linear combination of contact diagrams.

To prove \eqref{eq:vanishing_disc} and \eqref{eq:vanishing_z12}, we will make the simplifying assumption that the radial cutoff is taken to zero, $z_c\rightarrow 0$.\footnote{In the absence of a radial cutoff, we will regularize IR-divergent Witten diagrams using dimensional regularization \cite{Bzowski:2016kni}.}
If $\nu$ is real, then the $z_c\rightarrow 0$ limit of \eqref{eq:zc_0_GBB} can be written as \cite{Liu:1998ty},
\begin{align}
\mathcal{G}^{\text{AdS}}_{\nu}(k,z_1,z_2)=-i(z_1z_2)^{\frac{d}{2}}\int\limits_{0}^{\infty}dp \hspace{.05cm} p\frac{J_{\nu}(pz_1)J_{\nu}(pz_2)}{p^2+k^2-i\epsilon}.\label{eq:zc_0_GBB}
\end{align}
The identity \eqref{eq:zc_0_GBB} is the K\"all\'en-Lehmann spectral representation for the bulk-to-bulk propagator in the Poincar\'e patch.
Using \eqref{eq:zc_0_GBB} in \eqref{eq:diff_Witten_Feyn_disp} we find,
\begin{align}
A^{\text{Feyn}}_\O(s_-,z_1,z_2)-A^{\text{disp}}_\O(s_-,z_1,z_2)&=-i(z_1z_2)^{\frac{d}{2}}\int\limits_{0}^{\infty}\frac{dp}{p^2-s_- -i\epsilon}p\left(s_-^m-p^{2m}\right)J_{\nu}(pz_1)J_{\nu}(pz_2)
\nonumber \\
&=i(z_1z_2)^{\frac{d}{2}}\int\limits_{0}^{\infty}dp\hspace{.05cm}p \hspace{.05cm}\text{Poly}'(s_-,p^2)J_{\nu}(pz_1)J_{\nu}(pz_2),\label{eq:diff_amp_Pt1}
\end{align}
where $\text{Poly}'(s_-,p^2)$ is defined by,
\begin{align}
s_-^m-p^{2m}=(s_--p^2)\text{Poly}'(s_-,p^2).
\end{align}
Here we used a prime to distinguish this polynomial from the one that appears in the Witten diagram \eqref{eq:AdS_exchange_diagramV2}. 
From \eqref{eq:diff_amp_Pt1} we see that the difference between the two amputated diagrams has a vanishing $s$-channel discontinuity.
This proves the first condition, \eqref{eq:vanishing_disc}.

To prove \eqref{eq:vanishing_z12} we use the following completeness relation,
\begin{align}
\int\limits_0^{\infty}dp \hspace{.05cm} pJ_{\nu}(pz_1)J_{\nu}(pz_2)=\frac{\delta(z_1-z_2)}{z_1}.\label{eq:Bessel_Completness}
\end{align}
In order to prove that \eqref{eq:diff_amp_Pt1} only has support at $z_1=z_2$, we need to extend \eqref{eq:Bessel_Completness} by allowing for general, odd powers of $p$ in the integrand.
To do this, we can use the identity,
\begin{align}
\partial_{z} \big(z^{\nu}J_{\nu}(pz)\big)=p z^{\nu}J_{\nu-1}(pz),
\end{align}
to increase the power of $p$ in the left hand side of \eqref{eq:Bessel_Completness} by even integers.
For example,
\begin{align}
\int\limits_0^{\infty}dp \hspace{.05cm} p^3J_{\nu}(pz_1)J_{\nu}(pz_2)&=(z_1z_2)^{-(\nu+1)}\partial_{z_1}\partial_{z_2}\int\limits_0^{\infty}dp \hspace{.05cm} p (z_1z_2)^{\nu+1}J_{\nu+1}(pz_1)J_{\nu+1}(pz_2)
\nonumber
\\
&=(z_1z_2)^{-(\nu+1)}\partial_{z_1}\partial_{z_2}\left[ (z_1z_2)^{\nu+1/2}\delta(z_1-z_2)\right].\end{align}
We can repeat this procedure iteratively to prove that
\begin{align}
\int\limits_0^{\infty}dp \hspace{.05cm} p^{2m+1}J_{\nu}(pz_1)J_{\nu}(pz_2)=0, \quad \text{ if } \ z_1\neq z_2,
\end{align}
for arbitrary $m\in\mathbb{Z}_{\geq0}$.
This proves the second condition, \eqref{eq:vanishing_z12}, or that the difference between the dispersion formula result and the Feynman diagram is proportional to $\delta(z_1-z_2)$ and its derivatives.
This completes the proof that \eqref{eq:diff_Witten_Feyn_disp} is equal to a linear combination of contact Witten diagrams,\footnote{The sum over contact diagrams is finite if we assume that the two interaction vertices $V_{L,R}$ each involve a finite number of derivatives.}
\begin{align}
W^{\text{Feyn}}_{\O}(k_a,s_-,t_-)=W^{\text{disp}}_{\O}(k_a,s_-,t_-)+(\text{contact diagrams}).\label{eq:AdS_ambiguities}
\end{align}

After analytic continuation in $k$, \eqref{eq:AdS_ambiguities} proves our earlier claim for dS wavefunction coefficients, \eqref{eq:ambiguities}, when the exchanged dS field is light ($m^{2}<9/4$ in dS$_4$).
Here we are restricted to light fields because we need to assume that $\nu$ is real in order to use \eqref{eq:zc_0_GBB}.
For heavy fields in dS, or tachyons in AdS, $\nu$ is imaginary and there are additional terms when we take the $z_c\rightarrow 0$ limit of the bulk-to-bulk propagator, \eqref{eq:GBBAdS}.
In Appendix \ref{app:dispersion} we give an alternative proof of \eqref{eq:AdS_ambiguities} that does not require removing the radial cutoff and can be used to prove \eqref{eq:ambiguities} for heavy dS fields.

\section{Discussion}
\label{sec:Discussion}
In this work we developed unitarity methods for wavefunction coefficients in inflationary spacetimes.
In Section \ref{sec:dSWavefunction} we explained how to derive tree-level cutting rules and dispersion formulas for dS wavefunction coefficients by using the analyticity properties of the dS bulk-to-bulk propagator.
In Section \ref{sec:dSExchange} we used the cutting equation \eqref{eq:disc_gen_4pt} and dispersion formula \eqref{eq:ambiguities} to study generic, boost-breaking exchange diagrams that appear in the EFT of inflation.
In Section \ref{sec:AdSPerspective} we showed that the dS unitarity methods studied here can be understood as an analytic continuation of analogous methods in AdS.
Specifically, the bulk dS and AdS identities used in this work follow from the definition of the AdS propagator in terms of a time-ordered two-point function, or equivalently from the K\"all\'en-Lehmann spectral representation.
We did not need to assume that the bulk theory preserved boost symmetry, or equivalently that the boundary correlators were invariant under special conformal transformations.
We did assume that the boundary correlators are scale invariant because primordial fluctuations are known to be approximately scale invariant \cite{Planck:2019kim}.
However, this assumption was not required in principle.
 
There are many natural open questions to consider on how to extend our results to one loop and higher.
For example, are there cutting rules for $\disc_{s}\psi$ at higher loops?
In \cite{Meltzer:2020qbr} it was shown that the discontinuity of higher-loop AdS Witten diagrams can be computed using a set of cutting rules which are closely related to the flat space cutting rules \cite{Cutkosky:1960sp,Veltman:1963th}.
However, the AdS cutting rules involve $\theta$-functions with respect to the intermediate Lorentzian momenta and therefore they cannot be analytically continued to dS.
One resolution to this problem is to use unitarity to compute higher-loop, AdS Witten diagrams and then analytically continue the final result to dS.
While this does give a method to compute higher-loop wavefunction coefficients, including in boost-breaking theories, it would be interesting to understand if $\disc_{s}\psi$ has a direct connection to dS unitarity beyond tree level.
In unitary local CFTs, the momentum-space discontinuity of a time-ordered four-point function computes an advanced double-commutator \cite{Polyakov:1974gs,Meltzer:2020qbr,Meltzer:2021bmb}, which is the same out-of-time-ordered correlator that appears in the Lorentzian CFT inversion and dispersion formulas \cite{Caron-Huot:2017vep,Carmi:2019cub}.
It is then tempting to conjecture that $\disc_{s}\psi$, or perhaps the discontinuity of an in-in correlator, plays a similar role in cosmology.

Recently, in \cite{Melville:2021lst,Baumann:2021fxj} a set of dS loop-level cutting rules were derived for a different discontinuity than the one studied here.
At tree level, the dS cutting rules of \cite{Melville:2021lst,Baumann:2021fxj} and the AdS cutting rules of \cite{Meltzer:2020qbr} have the same effect: in both cases a cut bulk-to-bulk propagator factorizes in terms of a product of bulk-to-boundary propagators.
However, the two sets of cutting rules start to differ at one loop.
In AdS each cut factorizes a loop-level diagram into two disjoint halves while in dS one must include cuts that do not completely factorize the diagram.\footnote{A similar set of non-factorizing cuts appear in the Feynman tree theorem \cite{Feynman:1963ax,CaronHuot:2010zt}. See Appendix C of \cite{Meltzer:2020qbr} for an AdS generalization.}
It would be interesting to understand how the two sets of cutting rules might be related and more generally what the dS cutting rules can teach us about AdS/CFT and vice versa.

One can also consider extending the methods studied here to other spacetimes and to correlation functions of spinning operators.
For example, it should be possible to extend this work to more general FLRW spacetimes using the domain-wall/cosmology correspondence \cite{McFadden:2009fg}.
It would also be interesting to study correlation functions involving gauge bosons and gravitons and to understand their analytic structure in more detail at four points and higher.
Correlators of gauge bosons and gravitons are particularly interesting because they obey Ward-Takahashi identities, which imply that we need multiple exchange channels and specific contact diagrams to obtain a consistent correlator.
However, the dispersion formula studied here only reconstructs the full correlator if it vanishes in the momentum space Regge limit.
If not, it will generically only capture two out of the three exchange channels (specifically the $s$- and $u$-channels) and will have ambiguities associated to contact diagrams.
It would be interesting to understand the space of dS correlators which vanish in the Regge limit and are completely reconstructible from the dispersion formula studied here.\footnote{Alternatively, for Yang-Mills theory or Einstein gravity in pure (A)dS one could use the BCFW-like recursion relations developed in \cite{Raju:2010by,Raju:2011mp,Raju:2012zr,Raju:2012zs} to compute the full correlator.}
In addition, it would be interesting to apply the unitarity methods studied here to interactions which are not manifestly local.

Finally, we have used cutting rules and dispersion formulas to simplify dS perturbation theory.
In flat space \cite{Adams:2006sv} and AdS \cite{Hartman:2015lfa} one can go further and use dispersion formulas to derive rigorous bounds on the space of low-energy EFTs.\footnote{For recent works on bounding EFTs in flat space and AdS see \cite{Arkani-Hamed:2020blm,Chiang:2021ziz,Tolley:2020gtv,Caron-Huot:2020cmc,Bellazzini:2020cot,Sinha:2020win,Caron-Huot:2021rmr} and \cite{Penedones:2019tng,Kundu:2021qpi,Caron-Huot:2021enk}, respectively.}
It would be interesting to understand whether our methods also hold non-perturbatively and can be used to constrain models of inflation in dS using causality and unitarity, perhaps along the lines of \cite{Baumann:2015nta,Grall:2021xxm}.
This would give a method to chart out the space of consistent inflationary models and begin to understand their UV completions.
To derive rigorous bounds, it may also be useful to develop a non-perturbative bootstrap for the in-in correlators, as these are the physical observables measured in experiment. 

In this work we have explained how tree-level, $n$-point wavefunction coefficients in quasi-dS spacetimes can be bootstrapped from lower-point coefficients.
Both the dS unitarity methods studied here and the analogous AdS methods studied in \cite{Meltzer:2020qbr,Meltzer:2021bmb} are based on the unitarity methods developed in the original S-matrix bootstrap \cite{Eden:1966dnq}.
We expect that further investigations of unitarity and causality in approximate (A)dS spacetimes will teach us more about both holography in realistic spacetimes and the structure of non-Gaussianities in phenomenologically interesting models of the early universe.

\section*{Acknowledgements}
We thank Sadra Jazayeri, Savan Kharel, Junyu Liu, Enrico Pajer, Julio Parra-Martinez, David Simmons-Duffin, and Allic Sivaramakrishnan for discussions. The research of DM is supported by the Walter Burke Institute for Theoretical Physics and the Sherman Fairchild Foundation. This material is based upon work supported by the U.S. Department of Energy, Office of Science, Office of High Energy Physics, under Award Number DE-SC0011632.

\appendix
\section{Four-Point Diagrams}
\label{app:Four_Point}
\subsection{Quartic Interactions}
\label{app:quartic_interactions}
In this appendix we compute quartic, contact diagrams in dS for a massless scalar $\f$ ($\nu_{\f}=3/2$ or $\Delta_{\f}=3$),
\begin{align}
	\psi_{\text{contact}}(k_a,s,t)=\begin{aligned}
		\includegraphics[scale=.35]{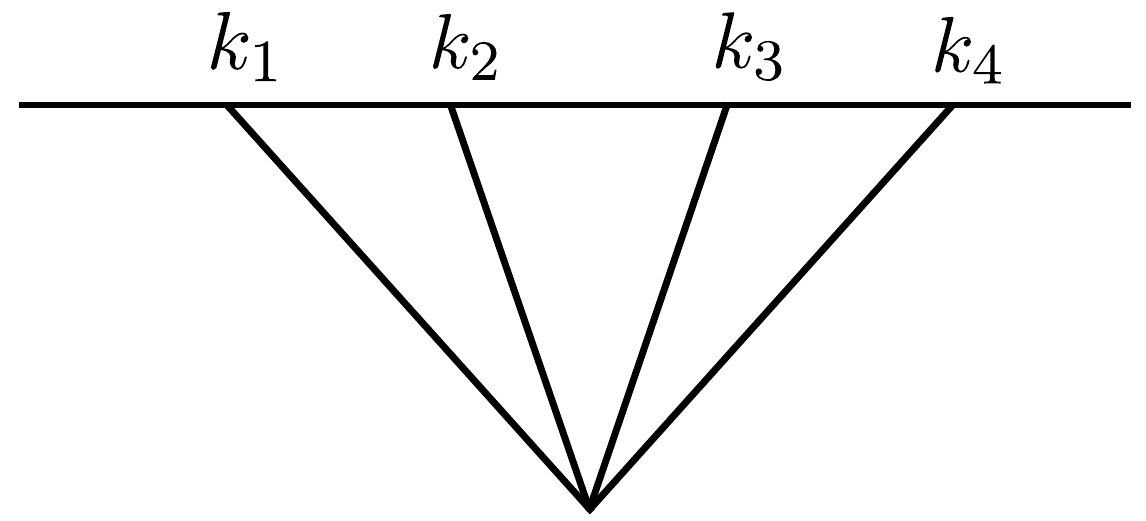}
	\end{aligned},
\end{align}
corresponding to the following interactions
\begin{align}
\mathcal{L}_{\text{quartic}}\supset \f'^4, \quad a(\eta)\f\f'^3,\quad a^{2}(\eta)\f^2\f'^2,\quad a^{2}(\eta)\f^2(\partial_i\f)^2.\label{eq:gen_quartic_int}
\end{align}
The scale factor in dS is $a(\eta)=-1/(H\eta)$. Here we continue to set $H=1$.
In \eqref{eq:gen_quartic_int} we have combined factors of $a(\eta)$ that come from the measure term, $\sqrt{-g}$, and those that come from the interactions themselves.
For IR finite contact diagrams we will set $\eta_0=0$.

The wavefunction coefficients are given by,
\begin{align}
\psi^{\f'^4}_{4}&=i\int\limits_{-\infty}^{0}d\eta\hspace{.05cm} a(\eta)\mathcal{K}'_{\f}(k_1,\eta)\mathcal{K}'_{\f}(k_2,\eta)\mathcal{K}'_{\f}(k_3,\eta)\mathcal{K}'_{\f}(k_4,\eta)
\nonumber \\
&=\frac{24 k_1^2 k_2^2 k_3^2 k_4^2}{(k_1+k_2+k_3+k_4)^5},
\end{align}
\begin{align}
\psi^{\f\f'^3}_{4}&=i\int\limits_{-\infty}^{0}d\eta\hspace{.05cm} a^{2}(\eta)\mathcal{K}'_{\f}(k_1,\eta)\mathcal{K}'_{\f}(k_2,\eta)\mathcal{K}'_{\f}(k_3,\eta)\mathcal{K}_{\f}(k_4,\eta)+\text{(symm)}
\nonumber
\\
&=\frac{2 k_1^2 k_2^2 k_3^2 (k_1+k_2+k_3+4 k_4)}{(k_1+k_2+k_3+k_4)^4}+\text{(symm)},
\\
\psi^{\f^2\f'^2}_{4}&=i\int\limits_{-\infty}^{0}d\eta\hspace{.05cm} a^{2}(\eta)\mathcal{K}'_{\f}(k_1,\eta)\mathcal{K}'_{\f}(k_2,\eta)\mathcal{K}_{\f}(k_3,\eta)\mathcal{K}_{\f}(k_4,\eta)+\text{(symm)}
\nonumber \\
&=\frac{k_1^2 k_2^2 \left(k_1^2+2 k_1 k_2+3 k_1 (k_3+k_4)+k_2^2+3 k_2 (k_3+k_4)+2 \left(k_3^2+3 k_3 k_4+k_4^2\right)\right)}{(k_1+k_2+k_3+k_4)^3}
\nonumber
\\
&\hspace{4.5in}+\text{(symm)},
 \\
\psi^{\f^2(\partial_i \f)^2}_{4}& = -i\int\limits_{-\infty}^{\eta_0}d\eta\hspace{.05cm} a^{2}(\eta)\mathcal{K}_{\f}(k_1,\eta)\mathcal{K}_{\f}(k_2,\eta)\mathcal{K}_{\f}(k_3,\eta)\mathcal{K}_{\f}(k_4,\eta)(\mathbf{k}_1\cdot\mathbf{k}_2+\text{symm})
\nonumber
\\
&=\frac{\sum_ak_a^{2}}{2(k_1+k_2+k_3+k_4)^3}\bigg(6k_1k_2k_3k_4+\sum\limits_{a=1}^{4}k_a^4+3\sum\limits_{ a\neq b}k_a^3k_b+4\sum\limits_{a<b}^{4}k_a^2k_b^2+6\sum\limits_{\substack{a\neq b,c \\ j<l}}k_a^2k_bk_c\bigg)
\nonumber
\\
&
\hspace{4.5in}-i\frac{\sum_ak_a^{2}}{\eta_0},\label{eq:divquartic}
\end{align}
where we summed over symmetrizations where relevant.
The last diagram, $\psi^{\f^2(\partial_i \f)^2}_{4}$, is IR divergent but can be regularized by subtracting off the local, power-law divergence,
\begin{align}
\psi^{\f^2(\partial_i \f)^2}_{4,\text{reg}}=\psi^{\f^2(\partial \f)^2}_{4}+i\frac{\sum_ak_a^{2}}{\eta_0}.
\end{align}

\subsection{Exchange Diagrams}
\label{sec:explicit_results}
In this appendix we give the full results for four-point exchange diagrams computed using the dispersion formula.
To simplify the notation, we will set all overall couplings to one.
\subsubsection*{$\boldsymbol{\f\f'^2\times\f\f'^2}$}
The exchange diagram constructed from two copies of the $\f\f'^2$ vertex is,
\begin{footnotesize}
\begin{align}
\psi^{\f\f'^2}_{4,\text{disp}}=&\frac{s^{3/2}}{\left((k_1+k_2)^2-s\right)^2 \left((k_3+k_4)^2-s\right)^2} \left(k_1^4+4 k_1^3 k_2-s \left(k_1^2+k_2^2\right)+4 k_1^2 k_2^2+4 k_1 k_2^3+k_2^4\right)
\nonumber
\\
&\times \left(-k_3^4-4 k_3^3 k_4+k_3^2 \left(s-4 k_4^2\right)-4 k_3 k_4^3-k_4^4+k_4^2 s\right)
\nonumber \\
&+\bigg\{\frac{1}{\left((k_1+k_2)^2-s\right)^2 (k_1+k_2-k_3-k_4)^3 (k_1+k_2+k_3+k_4)^3}
\nonumber
\\
&\hspace{.25in}\bigg[\left((k_1+k_2)^2-s\right) \big(2 k_1 k_2 \left(k_1^2+k_1 k_2+k_2^2\right) (k_1+k_2)^3 \left(k_3^2+k_4^2\right) (k_1+k_2-k_3-k_4) 
\nonumber
\\
&\hspace{.35in}(k_1+k_2+k_3+k_4)+\left(-(k_1+k_2)^2 \left(k_3^2+k_4^2\right)+k_3^4+4 k_3^3 k_4+4 k_3^2 k_4^2+4 k_3 k_4^3+k_4^4\right) 
\nonumber
\\
&\hspace{.35in}\big(\left(k_1^2+k_2^2\right) (k_1+k_2)^3 (k_1+k_2-k_3-k_4) (k_1+k_2+k_3+k_4)+k_1 k_2 \left(k_1^2+k_1 k_2+k_2^2\right)
\nonumber
\\
&\hspace{.35in} (k_1+k_2) \left((k_1+k_2)^2+3 (k_3+k_4)^2\right)\big)\big)+2 k_1 k_2 \left(k_1^2+k_1 k_2+k_2^2\right) (k_1+k_2)^3 (k_1+k_2-k_3-k_4)
\nonumber
\\
& \hspace{.35in} (k_1+k_2+k_3+k_4) \left(-(k_1+k_2)^2 \left(k_3^2+k_4^2\right)+k_3^4+4 k_3^3 k_4+4 k_3^2 k_4^2+4 k_3 k_4^3+k_4^4\right)\bigg]
\nonumber
\\
&\hspace{3.75in}+(k_1,k_2)\leftrightarrow(k_3,k_4)\bigg\}.
\end{align}
\end{footnotesize}
Despite appearances, this expression is regular in the folded limits $s\rightarrow \pm (k_1+k_2), \pm(k_3+k_4)$ as well as in the limit $k_1+k_2-k_3-k_4\rightarrow0$.

\subsubsection*{$\boldsymbol{\f'^3\times\f'^3}$}
The exchange diagram made from two $\f'^3$ vertices was given previously in \eqref{eq:finalEFT1squared}.
\subsubsection*{$\boldsymbol{\f'^3\times\f'(\partial_i\f)^2}$}
Here we study $s$-channel exchange diagrams constructed from a $\f'^3$ and $\f'(\partial_i\f)^2$ vertex.
There are two such diagrams, depending on which side we place each vertex.
The exchange diagram where the momenta $(\mathbf{k}_1,\mathbf{k}_2)$ flow into the $\f'^3$ vertex and the momenta $(\mathbf{k}_3,\mathbf{k}_4)$ flow into the $\f'(\partial_i\f)^2$ vertex is,

\begin{scriptsize}
\begin{align}
\hspace{-.35in}\psi^{\f'^3\times \f'(\partial_i\f)^2}_{4,\text{disp}}=&\frac{k_1^2 k_2^2 s^{3/2} \left(3 (k_1+k_2)^2+s\right) }{\left((k_1+k_2)^2-s\right)^3 \left((k_3+k_4)^2-s\right)^3}
\bigg(s^2 \left(7 k_3^2+6 k_3 k_4+7 k_4^2\right)-s \left(11 k_3^4+36 k_3^3 k_4+54 k_3^2 k_4^2+36 k_3 k_4^3+11 k_4^4\right)
\nonumber
\\
&\hspace{2.2in}+(k_3+k_4)^2 \left(5 k_3^4+20 k_3^3 k_4-14 k_3^2 k_4^2+20 k_3 k_4^3+5 k_4^4\right)-s^3\bigg)
\nonumber
\\
&-\bigg\{
\frac{ k_1^2 k_2^2 (k_1+k_2)}{(k_1+k_2-k_3-k_4)^5 (k_1+k_2+k_3+k_4)^5 \left((k_1+k_2)^2-s\right)^3}
\nonumber
\\
&\hspace{.3in}\bigg((k_1+k_2)^2 (k_1+k_2-k_3-k_4)^2 \big((k_1+k_2)^6-7 k_3^2 (k_1+k_2)^4-7 k_4^2 (k_1+k_2)^4-6 k_3 k_4 (k_1+k_2)^4
\nonumber
\\
&\hspace{.3in}+11 k_3^4 (k_1+k_2)^2+11 k_4^4 (k_1+k_2)^2+36 k_3 k_4^3 (k_1+k_2)^2+54 k_3^2 k_4^2 (k_1+k_2)^2+36 k_3^3 k_4 (k_1+k_2)^2
\nonumber
\\
&\hspace{.3in}-5 k_3^6-5 k_4^6-30 k_3 k_4^5-31 k_3^2 k_4^4-12 k_3^3 k_4^3-31 k_3^4 k_4^2-30 k_3^5 k_4\big) \left(3 (k_1+k_2)^2+s\right) (k_1+k_2+k_3+k_4)^2
\nonumber
\\
&\hspace{.3in}-(k_1+k_2-k_3-k_4) \big(4 (k_1+k_2)^4 (k_1+k_2-k_3-k_4) (k_1+k_2+k_3+k_4) (3 (k_1+k_2)^4
\nonumber
\\
&\hspace{.3in}
-14 k_3^2 (k_1+k_2)^2-14 k_4^2 (k_1+k_2)^2-12 k_3 k_4 (k_1+k_2)^2+11 k_3^4+11 k_4^4+36 k_3 k_4^3+54 k_3^2 k_4^2+36 k_3^3 k_4)
\nonumber
\\
&\hspace{.3in}
-6 (k_1+k_2)^2 \big((k_1+k_2)^6-7 k_3^2 (k_1+k_2)^4-7 k_4^2 (k_1+k_2)^4-6 k_3 k_4 (k_1+k_2)^4+11 k_3^4 (k_1+k_2)^2
\nonumber
\\
&\hspace{.3in}
+11 k_4^4 (k_1+k_2)^2+36 k_3 k_4^3 (k_1+k_2)^2+54 k_3^2 k_4^2 (k_1+k_2)^2+36 k_3^3 k_4 (k_1+k_2)^2-5 k_3^6-5 k_4^6-30 k_3 k_4^5
\nonumber
\\
&\hspace{.3in}
-31 k_3^2 k_4^4-12 k_3^3 k_4^3-31 k_3^4 k_4^2-30 k_3^5 k_4\big) \left((k_1+k_2)^2+(k_3+k_4)^2\right)\big) \left((k_1+k_2)^2-s\right) (k_1+k_2+k_3+k_4)
\nonumber
\\
&\hspace{.3in}
+\bigg((k_1+k_2-k_3-k_4)^2 (k_1+k_2+k_3+k_4)^2 \big(15 (k_1+k_2)^4-42 k_3^2 (k_1+k_2)^2-42 k_4^2 (k_1+k_2)^2
\nonumber
\\
&\hspace{.3in}
-36 k_3 k_4 (k_1+k_2)^2+11 k_3^4+11 k_4^4+36 k_3 k_4^3+54 k_3^2 k_4^2+36 k_3^3 k_4\big) (k_1+k_2)^2
\nonumber
\\
&\hspace{.3in}
-6 (k_1+k_2-k_3-k_4) (k_1+k_2+k_3+k_4) \big(3 (k_1+k_2)^4-14 k_3^2 (k_1+k_2)^2-14 k_4^2 (k_1+k_2)^2
\nonumber
\\
&\hspace{.3in}-12 k_3 k_4 (k_1+k_2)^2+11 k_3^4+11 k_4^4+36 k_3 k_4^3+54 k_3^2 k_4^2+36 k_3^3 k_4\big) \left((k_1+k_2)^2+(k_3+k_4)^2\right) (k_1+k_2)^2
\nonumber
\\
&\hspace{.3in}
+3 \big((k_1+k_2)^6-7 k_3^2 (k_1+k_2)^4
-7 k_4^2 (k_1+k_2)^4-6 k_3 k_4 (k_1+k_2)^4+11 k_3^4 (k_1+k_2)^2+11 k_4^4 (k_1+k_2)^2
\nonumber
\\
&\hspace{.3in}
+36 k_3 k_4^3 (k_1+k_2)^2+54 k_3^2 k_4^2 (k_1+k_2)^2+36 k_3^3 k_4 (k_1+k_2)^2-5 k_3^6-5 k_4^6-30 k_3 k_4^5
\nonumber
\\
&\hspace{.3in}
-31 k_3^2 k_4^4-12 k_3^3 k_4^3-31 k_3^4 k_4^2-30 k_3^5 k_4\big) (2 k_1^4+8 k_2 k_1^3+\left(12 k_2^2+5 (k_3+k_4)^2\right) k_1^2
\nonumber
\\
&\hspace{.3in}
+2 k_2 \left(4 k_2^2+5 (k_3+k_4)^2\right) k_1+2 k_2^4+(k_3+k_4)^4+5 k_2^2 (k_3+k_4)^2)\bigg) \left((k_1+k_2)^2-s\right)^2\bigg)
\nonumber
\\
&\hspace{4in}+(k_1,k_2)\leftrightarrow(k_3,k_4)\bigg\}.
\end{align}
\end{scriptsize}
The exchange diagram with the two vertices flipped is,
\begin{align}
\psi^{\f'(\partial_i\f)^2\times \f'^3}_{4,\text{disp}}=\psi^{\f'^3\times \f'(\partial_i\f)^2}_{4,\text{disp}}\bigg|_{(k_1,k_2)\leftrightarrow (k_3,k_4)}.
\end{align}
As a consistency check, one can show that each diagram is regular in the folded limits
\\
 $s\rightarrow \pm(k_1+k_2)^2, \ \pm(k_3+k_4)^2$ and $k_1+k_2-k_3-k_4\rightarrow0$.
In addition, each diagram satisfies the manifestly local test \cite{Jazayeri:2021fvk},
\begin{align}
\partial_{k_a}\psi^{\f'^3\times \f'(\partial_i\f)^2}_{4,\text{disp}}\bigg|_{k_a=0}=0.
\end{align}

\subsubsection*{$\boldsymbol{\f'(\partial_i\f)^2\times\f'(\partial_i\f)^2}$}
Here we give the dispersive representation for the exchange diagram constructed from two $\f'(\partial_i\f)^2$ vertices.
We find the following expression,
\begin{scriptsize}
\begin{align}
\hspace{-.25in}\psi^{\f'(\partial_i\f)^2}_{4,\text{disp}}=&
-\frac{s^{3/2}}{4 \left((k_1+k_2)^2-s\right)^3 \left((k_3+k_4)^2-s\right)^3}
\nonumber
\\
&
 \big(s^2 \left(7 k_1^2+6 k_1 k_2+7 k_2^2\right)-s \left(11 k_1^4+36 k_1^3 k_2+54 k_1^2 k_2^2+36 k_1 k_2^3+11 k_2^4\right)
 \nonumber
\\
&
\hspace{.1in} +(k_1+k_2)^2 \left(5 k_1^4+20 k_1^3 k_2-14 k_1^2 k_2^2+20 k_1 k_2^3+5 k_2^4\right)-s^3\big) 
 \nonumber
\\
&
\big(s^2 \left(7 k_3^2+6 k_3 k_4+7 k_4^2\right)-s \left(11 k_3^4+36 k_3^3 k_4+54 k_3^2 k_4^2+36 k_3 k_4^3+11 k_4^4\right)
 \nonumber
\\
&
\hspace{.1in} 
+(k_3+k_4)^2 \left(5 k_3^4+20 k_3^3 k_4-14 k_3^2 k_4^2+20 k_3 k_4^3+5 k_4^4\right)-s^3\big)
\nonumber
\\
&
-\bigg\{
\frac{1}{4 (k_1+k_2) (k_1+k_2-k_3-k_4)^5 (k_1+k_2+k_3+k_4)^5 \left(k_1^2+2 k_2 k_1+k_2^2-s\right)^3}
\nonumber
\\
&
\hspace{.2in}
\bigg(
-8 k_1 k_2 (k_1+k_2-k_3-k_4) (k_1+k_2+k_3+k_4) \big(6 k_1 k_2 (k_1+k_2-k_3-k_4) (k_1+k_2+k_3+k_4)
\nonumber
\\
&
\hspace{.2in}
 \left(3 (k_1+k_2)^4-2 \left(7 k_3^2+6 k_4 k_3+7 k_4^2\right) (k_1+k_2)^2+11 k_3^4+11 k_4^4+36 k_3 k_4^3+54 k_3^2 k_4^2+36 k_3^3 k_4\right) (k_1+k_2)^2
 \nonumber
\\
&
\hspace{.2in}
 +\big((k_1+k_2)^6-\left(7 k_3^2+6 k_4 k_3+7 k_4^2\right) (k_1+k_2)^4+\left(11 k_3^4+36 k_4 k_3^3+54 k_4^2 k_3^2+36 k_4^3 k_3+11 k_4^4\right) (k_1+k_2)^2
  \nonumber
\\
&
\hspace{.2in}
-(k_3+k_4)^2 \left(5 k_3^4+20 k_4 k_3^3-14 k_4^2 k_3^2+20 k_4^3 k_3+5 k_4^4\right)\big) (k_1^4-6 k_2 k_1^3-\left(14 k_2^2+(k_3+k_4)^2\right) k_1^2
\nonumber
\\
&
\hspace{.2in}
-2 k_2 \left(3 k_2^2+5 (k_3+k_4)^2\right) k_1+k_2^2 \left(k_2^2-(k_3+k_4)^2\right))\big) \left(k_1^2+2 k_2 k_1+k_2^2-s\right) (k_1+k_2)^4
\nonumber
\\
&
\hspace{.2in}
+12 k_1^2 k_2^2 (k_1+k_2-k_3-k_4)^2 (k_1+k_2+k_3+k_4)^2 \big((k_1+k_2)^6-\left(7 k_3^2+6 k_4 k_3+7 k_4^2\right) (k_1+k_2)^4
\nonumber
\\
&
\hspace{.2in}
+\left(11 k_3^4+36 k_4 k_3^3+54 k_4^2 k_3^2+36 k_4^3 k_3+11 k_4^4\right) (k_1+k_2)^2-(k_3+k_4)^2 \left(5 k_3^4+20 k_4 k_3^3-14 k_4^2 k_3^2+20 k_4^3 k_3+5 k_4^4\right)\big) 
\nonumber
\\
&
\hspace{.2in}
\left(3 k_1^2+6 k_2 k_1+3 k_2^2+s\right) (k_1+k_2)^4+\big(12 k_1^2 k_2^2 (k_1+k_2-k_3-k_4)^2 (k_1+k_2+k_3+k_4)^2 (15 (k_1+k_2)^4
\nonumber
\\
&
\hspace{.2in}
-6 \left(7 k_3^2+6 k_4 k_3+7 k_4^2\right) (k_1+k_2)^2+11 k_3^4+11 k_4^4+36 k_3 k_4^3+54 k_3^2 k_4^2+36 k_3^3 k_4) (k_1+k_2)^4
\nonumber
\\
&
\hspace{.2in}
+8 k_1 k_2 (k_1+k_2-k_3-k_4) (k_1+k_2+k_3+k_4) \big(3 (k_1+k_2)^4-2 \left(7 k_3^2+6 k_4 k_3+7 k_4^2\right) (k_1+k_2)^2+11 k_3^4+11 k_4^4
\nonumber
\\
&
\hspace{.2in}
+36 k_3 k_4^3+54 k_3^2 k_4^2+36 k_3^3 k_4\big) \big(k_1^4-6 k_2 k_1^3-\left(14 k_2^2+(k_3+k_4)^2\right) k_1^2-2 k_2 \left(3 k_2^2+5 (k_3+k_4)^2\right) k_1
\nonumber
\\
&
\hspace{.2in}
+k_2^2 \left(k_2^2-(k_3+k_4)^2\right)\big) (k_1+k_2)^4-\big((k_1+k_2)^6-\left(7 k_3^2+6 k_4 k_3+7 k_4^2\right) (k_1+k_2)^4+(11 k_3^4+36 k_4 k_3^3
\nonumber
\\
&
\hspace{.2in}
+54 k_4^2 k_3^2+36 k_4^3 k_3+11 k_4^4) (k_1+k_2)^2-(k_3+k_4)^2 \left(5 k_3^4+20 k_4 k_3^3-14 k_4^2 k_3^2+20 k_4^3 k_3+5 k_4^4\right)\big) 
\nonumber
\\
&
\hspace{.2in}
\big(-18 k_1^2 (k_1+k_2) \big(5 k_1^4+20 k_2 k_1^3+10 \left(3 k_2^2+(k_3+k_4)^2\right) k_1^2+20 k_2 \left(k_2^2+(k_3+k_4)^2\right) k_1+5 k_2^4+(k_3+k_4)^4
\nonumber
\\
&
\hspace{.2in}
+10 k_2^2 (k_3+k_4)^2\big) k_2^2+k_1 \left(2 k_1^3+7 k_2 k_1^2+7 k_2^2 k_1+2 k_2^3\right) (k_1+k_2-k_3-k_4) (k_1+k_2+k_3+k_4)
\nonumber
\\
&
\hspace{.2in}
\left(7 k_1^2+14 k_2 k_1+7 k_2^2+5 (k_3+k_4)^2\right) k_2+(k_1+k_2) \left(4 k_1^4+6 k_2 k_1^3+3 k_2^2 k_1^2+6 k_2^3 k_1+4 k_2^4\right) (k_1+k_2-k_3-k_4)^2
\nonumber
\\
&
\hspace{.2in}
(k_1+k_2+k_3+k_4)^2\big) (k_1+k_2)\big) \left(k_1^2+2 k_2 k_1+k_2^2-s\right)^2
\bigg)
+(k_1,k_2)\leftrightarrow(k_3,k_4)
\bigg\}.
\label{eq:EFT2AppResult}
\end{align}
\end{scriptsize}
This expression is also finite in the folded limits and passes the manifestly local test,
\begin{align}
\partial_{k_a}\psi^{\f'(\partial_i\f)^2}_{4,\text{disp}}\bigg|_{k_a=0}=0.
\end{align}
Furthermore, as demonstrated by \eqref{eq:diff_quartic}, our result differs from the result found in \cite{Jazayeri:2021fvk} by a finite sum of contact diagrams.

\section{Regularizing Dispersion Integrals}
\label{sec:Reg_Dispersion}
In this appendix we discuss two different ways of regularizing divergent dispersion integrals in dS: either by cutting off the $\eta$ integral or by defining divergent integrals through analytic continuation. 
We will only discuss the exchange diagram constructed from two $\f'(\partial_i\f)^2$ vertices, although our analysis carries over to general diagrams.
We show that for this diagram, the difference between the two regularizations is given by a sum of quartic, contact diagrams.
\subsection*{Hard Cutoff}
To use the hard cutoff regularization, we need to compute the three-point function associated to $\f'(\partial_i\f)^2$ at non-zero $\eta_0$,
\begin{align}
\psi^{\f'(\partial_i\f)^2}_{3,\text{reg}}(k_1,k_2,k_3)=&\ ig_{2}\mathbf{k}_2\cdot\mathbf{k}_3\int\limits_{-\infty}^{\eta_0}\frac{d\eta}{\eta}\mathcal{K}'_{\f}(k_1,\eta)\mathcal{K}_{\f}(k_2,\eta)\mathcal{K}_{\f}(k_3,\eta)+\text{(symm)}
\nonumber
\\
=& \ \frac{k_1^2 \mathbf{k}_2\cdot\mathbf{k}_3 }{(\eta_0 k_1-i) (\eta_0 k_2-i) (\eta_0 k_3-i) (k_1+k_2+k_3)^3}
\nonumber
\\
&\bigg(i \left(k_1^2+3 k_1 (k_2+k_3)+2 \left(k_2^2+3 k_2 k_3+k_3^2\right)\right)
\nonumber
\\
&-\eta_0 (k_1+k_2+k_3) \left(k_1 (k_2+k_3)+k_2^2+4 k_2 k_3+k_3^2\right)-i \eta_0^2 k_2 k_3 (k_1+k_2+k_3)^2\bigg)
\nonumber
\\
&\hspace{4.3in}+\text{(symm)}
,\label{eq:three_pt_regularized}
\end{align}
where we sum over symmetrizations. 
We can now plug this result into the dispersion formula,
\begin{align}
\psi^{\f'(\partial_i\f)^2}_{4,\text{reg}}(k_a,s,t)=&\frac{1}{2\pi i}\int\limits_{0}^{\infty}\frac{ds'}{s'-s+i\epsilon}\disc_{s'}\psi^{\f'(\partial_i\f)^2}_{4,\text{reg}}(k_a,s',t)
\nonumber
\\
=&\frac{1}{2\pi i}\int\limits_{0}^{\infty}\frac{ds'}{s'-s+i\epsilon}P_{3/2}(\sqrt{s'})\disc_{s'}\psi^{\f'(\partial_i\f)^2}_{3,\text{reg}}(k_1,k_2,\sqrt{s'})\disc_{s'}\psi^{\f'(\partial_i\f)^2}_{3,\text{reg}}(\sqrt{s'},k_3,k_4).\label{eq:disp_app_EFT2}
\end{align}
In this case we find the discontinuity vanishes in the large $s$ limit,
\begin{align}
\disc_{s}\psi^{\f'(\partial_i\f)^2}_{4,\text{reg}}(k_a,s,t)\sim s^{-1} \quad \text{ for } \ s\gg1, 
\end{align}
and the dispersion integral \eqref{eq:disp_app_EFT2} converges. 
To evaluate the dispersion integral, we make the change of variables $s'=p^2$, extend the $p$-contour to $(-\infty,\infty)$, and then close the contour in the lower half-plane,
\begin{align}
\psi^{\f'(\partial_i\f)^2}_{4,\text{reg}}(k_a,s,t)&=\frac{1}{2\pi i}\int\limits_{-\infty}^{\infty}\frac{dp}{p^2-s+i\epsilon}p\hspace{.05cm}\disc_{p^2}\psi^{\f'(\partial_i\f)^2}_{4,\text{reg}}(k_a,p^2,t)
\nonumber
\\
=&-\left(\res\limits_{p=k_1+k_2}+\res\limits_{p=k_3+k_4}+\res\limits_{p=\sqrt{s-i\epsilon}}+\res\limits_{p=i/\eta_0 }\right)\frac{p}{p^2-s+i\epsilon}\disc_{p^2}\psi^{\f'(\partial_i\f)^2}_{4,\text{reg}}(k_a,p^2,t).\label{eq:App_five_res}
\end{align}
The new feature in comparison to the cases studied in Section \ref{sec:dSExchange} is that we now have a pole at $p=i/\eta_0$.
In the limit $\eta_0\rightarrow 0^{-}$, the first three poles give the same result as before \eqref{eq:EFT2AppResult},
\begin{align}
\psi^{\f'(\partial_i\f)^2}_{4,\text{disp}}(k_a,s,t)=-\lim\limits_{\eta_0\rightarrow0}\left(\res\limits_{p=k_1+k_2}+\res\limits_{p=k_3+k_4}+\res\limits_{p=\sqrt{s-i\epsilon}}\right)\frac{p}{p^2-s+i\epsilon}\disc_{p^2}\psi^{\f'(\partial_i\f)^2}_{4,\text{reg}}(k_a,p^2,t).
\end{align}
The difference between the two results, \eqref{eq:EFT2AppResult} and \eqref{eq:App_five_res}, then comes from the pole at $p=i/\eta_0$,
\begin{align}
\psi^{\f'(\partial_i\f)^2}_{4,\text{disp}}(k_a,s,t)-\psi^{\f'(\partial_i\f)^2}_{4,\text{reg}}(k_a,s,t)=&\res\limits_{p=i/\eta_0 }\frac{p}{p^2-s+i\epsilon}\disc_{p^2}\psi^{\f'(\partial_i\f)^2}_{4,\text{reg}}(k_a,p^2,t)
\nonumber
\\[5pt]
=&\frac{i}{8 \eta_0^3}-\frac{i \left(6 k_1^2+6 k_2^2+6 k_3^2+6 k_4^2-s\right)}{8 \eta_0}
\nonumber
\\
&\hspace{.66cm}-\frac{1}{4} \left(k_1^3+k_2^3+k_3^3+k_4^3\right) +O(\eta_0),\label{eq:diff_reg_APP}
\end{align}
where we have dropped terms that vanish in the limit $\eta_0\rightarrow0^{-}$.

The first two terms in \eqref{eq:diff_reg_APP} are analytic in all of the external momenta and correspond to ultra-local contact terms in position space, that is they only have support when all four external points are coincident, $\mathbf{x}_1=\mathbf{x}_2=\mathbf{x}_3=\mathbf{x}_4$.
These can be removed by adding local counterterms.
The third term is semi-local in position space since it is analytic in a subset of the momenta.
For example, $k_1^3$ is non-analytic in $\mathbf{k}_1$, but is analytic in $\mathbf{k}_{2,3}$.
The third term in \eqref{eq:diff_reg_APP} comes from the following combination of contact diagrams,
\begin{align}
\psi^{\f^2\f'^2}_{4}(k_a,s,t)-\psi^{\f^2(\partial_i\f)^2}_{4,\text{reg}}(k_a,s,t)=\frac{1}{2}(k_1^3+k_2^3+k_3^3+k_4^3).
\end{align}
Finally, we find that the difference between \eqref{eq:EFT2AppResult} and \eqref{eq:App_five_res} is a linear combination of contact diagrams,
\begin{align}
\psi^{\f'(\partial_i\f)^2}_{4,\text{disp}}(k_a,s,t)-\psi^{\f'(\partial_i\f)^2}_{4,\text{reg}}(k_a,s,t)=-\frac{1}{2}\left(\psi^{\f^2\f'^2}_{4}(k_a,s,t)-\psi^{\f^2(\partial_i\f)^2}_{4,\text{reg}}(k_a,s,t)\right).
\end{align}

\subsection*{Analytic Continuation}
In this section we explain how to define divergent dispersion integrals by analytic continuation.
Specifically, we will compute,
\begin{align}
\psi^{\f'(\partial_i\f)^2}_{4,\text{disp}}(k_a,s,t)=\frac{1}{2\pi i}\int\limits_{0}^{\infty}\frac{ds'}{s'-s+i\epsilon}\disc_{s'}\psi^{\f'(\partial_i\f)^2}_{4}(k_a,s',t).\label{eq:disp_int_EFT2_APP}
\end{align}
First we recall that if we use the unregulated three-point functions as input, then $\disc_{s}\psi^{\f'(\partial_i\f)^2}_{4}$ grows at large $s$, see \eqref{eq:growing_terms}.
For convenience we have copied the expression below with $g_2=1$,
\begin{align}
\lim\limits_{s\rightarrow\infty}\text{disc}_{s}\psi^{\f'(\partial_i\f)^2}_{4}(k_a,s,t)\approx \left(\frac{s^{3/2}}{2}-2s^{1/2}(k_1^2+k_2^2+k_3^2+k_4^2)\right).\label{eq:growing_terms_APP}
\end{align}
We can then define a subtracted integrand by,
\begin{align}
\mathcal{I}^{\f'(\partial_i\f)^2}_4(k_a,s,t)=\text{disc}_{s}\psi^{\f'(\partial_i\f)^2}_{4}(k_a,s,t)-\left(\frac{s^{3/2}}{2}-2s^{1/2}(k_1^2+k_2^2+k_3^2+k_4^2)\right).
\end{align}
Using $\mathcal{I}^{\f'(\partial_i\f)^2}_4$ we can rewrite the dispersion integral \eqref{eq:disp_int_EFT2_APP} as,
\begin{align}
\psi^{\f'(\partial_i\f)^2}_{4,\text{disp}}(k_a,s,t)=&\frac{1}{2\pi i}\int\limits_{0}^{\infty}\frac{ds'}{s'-s+i\epsilon}\mathcal{I}^{\f'(\partial_i\f)^2}_4(k_a,s',t)
\nonumber
\\
&+\frac{1}{2\pi i}\int\limits_{0}^{\infty}\frac{ds'}{s'-s+i\epsilon}\left(\frac{s'^{3/2}}{2}-2s'^{1/2}(k_1^2+k_2^2+k_3^2+k_4^2)\right),
\label{eq:app_reg_disp_AC}
\end{align}
where we have simply added and subtracted the power law terms.
The integral in the first line of \eqref{eq:app_reg_disp_AC} converges and can be computed in the standard way by making the change of variables $s'=p^2$ and evaluating the $p$ integral using Cauchy's theorem.
To define the second line we need to compute the dispersive transform of a pure power-law,
\begin{align}
S_{a}(s)=\frac{1}{2\pi i}\int\limits_{0}^{\infty}\frac{ds'}{s'-s+i\epsilon}s'^{a}=i\frac{e^{i \pi  a} }{2 \sin (\pi  a)}(s-i \epsilon)^a.\label{eq:disp_power_law}
\end{align}
The integral \eqref{eq:disp_power_law} only converges for $-1<a<0$, but we can define it for generic $a$ by analytic continuation.
Then the full dispersive integral is,
\begin{align}
\psi^{\f'(\partial_i\f)^2}_{4,\text{disp}}(k_a,s,t)=&\frac{1}{2\pi i}\int\limits_{0}^{\infty}\frac{ds'}{s'-s+i\epsilon}\mathcal{I}^{\f'(\partial_i\f)^2}_4(k_a,s',t)
\nonumber
\\
&+\left(\frac{1}{2}S_{3/2}(s)-2S_{1/2}(s)(k_1^2+k_2^2+k_3^2+k_4^2)\right).
\label{eq:app_reg_disp_ACV2}
\end{align}
In \eqref{eq:app_reg_disp_ACV2} both the first and second line are well-defined and this gives an expression whose $s$-channel discontinuity factorizes correctly.

We can also check that \eqref{eq:app_reg_disp_ACV2} agrees with what we would have found had we turned the original dispersion integral \eqref{eq:disp_int_EFT2_APP} into a contour integral in $p$ and dropped the arc at infinity.
To show this we use the following basic identity,
\begin{align}
-\res_{p=\sqrt{s-i\epsilon}}\frac{1}{p^{2}-s+i\epsilon}p^{2a+1}=-\frac{1}{2}(s-i\epsilon)^a,
\end{align}
where we included the overall minus sign because this pole is located in the lower half-plane.
If we set $a$ to be a half-integer, $a=n+1/2$ with $n\in\mathbb{Z}$, then this agrees with $S_{a}(s)$,
\begin{align}
S_{n+1/2}(s)=-\res_{p=\sqrt{s-i\epsilon}}\frac{1}{p^{2}-s+i\epsilon}p^{2(n+1)}.
\end{align}
Therefore, we could equivalently have found \eqref{eq:app_reg_disp_ACV2} by rewriting the dispersion integral as a contour integral and dropping the arc at infinity.
This trick is also valid for all exchange diagrams in the EFT of inflation which are constructed from rational three-point functions of the form \eqref{eq:threeptansatz}.

\section{Dispersion Formula Ambiguities}
\label{app:dispersion}
In this appendix we will give an alternative proof of \eqref{eq:AdS_ambiguities}, or that the dispersion formula reconstructs exchange Witten diagrams from their cuts, up to contact diagram ambiguities.
Our starting point is the identity \eqref{eq:diff_Witten_Feyn_disp}, which we reproduce below:
\begin{align}
W^{\text{Feyn}}_{\O}&(k_a,s_-,t_-)-W^{\text{disp}}_{\O}(k_a,s_-,t_-)
\nonumber
\\[3pt]
=&-\prod_ak^{n_a}_a \int\limits\frac{dz_1dz_2}{z_1^{d+1}z_2^{d+1}}
\mathcal{D}_{L,z_1}\mathcal{D}_{R,z_2}\left(\mathcal{K}^{\text{AdS}}_{\f}(k_1,z_1)\mathcal{K}^{\text{AdS}}_{\f}(k_2,z_1)\mathcal{K}^{\text{AdS}}_{\f}(k_3,z_2)\mathcal{K}^{\text{AdS}}_{\f}(k_4,z_2)\right)
\nonumber
\\
&\hspace{1.33in}\left(A^{\text{Feyn}}_\O(s_-,z_1,z_2)-A^{\text{disp}}_\O(s_-,z_1,z_2)\right),\label{eq:diff_Witten_Feyn_dispAPP}
\end{align}
where the amputated diagrams are defined by,
\begin{align}
A^{\text{Feyn}}_\O(s_-,z_1,z_2)&=s_-^m \hspace{.05cm}\mathcal{G}^{\text{AdS}}_{\O}(\sqrt{s_-},z_1,z_2),
\\
A^{\text{disp}}_\O(s_-,z_1,z_2)&=\frac{1}{2\pi i}\int\limits_{0}^{\infty}\frac{ds_{-}'}{s_{-}'-s_- -i\epsilon}s_-'^m \disc_{s_-} \mathcal{G}^{\text{AdS}}_{\O}\left(\sqrt{s_-'},z_1,z_2\right).\label{eq:APP_disp_amputated}
\end{align}
We will next Fourier transform each amputated diagram to position space.
We want to show that the difference between the amputated diagrams is localized at a single bulk point,
\begin{align}
A^{\text{Feyn}}_\O(\mathbf{x}_{12},z_1,z_2)-A^{\text{disp}}_\O(\mathbf{x}_{12},z_1,z_2)=0 \quad \text{ if } \ (\mathbf{x}_1,z_1)\neq (\mathbf{x}_2,z_2).\label{eq:APP_single_bulk}
\end{align}
The amputated Feynman diagram is given by,
\begin{align}
A^{\text{Feyn}}_\O(\mathbf{x}_{12},z_1,z_2)&=\partial_{\mathbf{x}_{12}}^{2m}\mathcal{G}_{\O}^{\text{AdS}}(\mathbf{x}_{12},z_1,z_2)
\nonumber
\\
&=\partial_{\mathbf{x}_{12}}^{2m}\<T[\Phi(x_1,z_1)\Phi(x_2,z_2)]\>_{\text{free}},
\end{align}
where we used the definition of the AdS propagator in terms of a time-ordered two-point function \eqref{eq:Def_G_Tord}.
Similarly the amputated, ``dispersive" diagram in position space is,
\begin{align}
A^{\text{disp}}_\O(\mathbf{x}_{12},z_1,z_2)=\theta(x_{12}^0)\partial_{\mathbf{x}_{12}}^{2m}\<\Phi(x_1,z_1)\Phi(x_2,z_2)\>_{\text{free}}+(1\leftrightarrow 2).\label{eq:disp_in_x_space}
\end{align}
To prove \eqref{eq:disp_in_x_space} we first Fourier transform \eqref{eq:APP_disp_amputated},
\begin{align}
A^{\text{disp}}_\O(\mathbf{x}_{12},z_1,z_2)=\frac{1}{2\pi i}\int \frac{d^{d}k}{(2\pi)^d}e^{-i\mathbf{k}\cdot \mathbf{x}_{12}}\int\limits_{0}^{\infty}\frac{ds'_-}{s'_- - s_- -i\epsilon}s_-'^m\disc_{s'_-}\mathcal{G}^{\text{AdS}}_{\nu}(k',z_1,z_2),
\end{align}
where $s_-'=-\mathbf{k}'^2$ and similarly for the unprimed variables.
Depending on the sign of $x_{12}^0$, we can either close the $k^0$ integral in the lower or upper half-plane.\footnote{There can be additional $\delta$-function terms localized at $\mathbf{x}_{12}=0$ which we will ignore here.}
The difference in how we close the contour can be accounted for by inserting different $\delta$- and $\theta$-functions in the integrand,
\begin{align}
A^{\text{disp}}_\O(\mathbf{x}_{12},z_1,z_2)&=\int \frac{d^{d}k}{(2\pi)^d}e^{-i\mathbf{k}\cdot \mathbf{x}_{12}}\int\limits_{0}^{\infty}ds'_-\hspace{.05cm}\delta(s'_--s_-)\left(\theta(x_{12}^0)\theta(-k^0)+\theta(x_{21}^0)\theta(k^0)\right)
\nonumber
\\
&\hspace{2.45in}s_-'^m\disc_{s'_-}\mathcal{G}^{\text{AdS}}_{\nu}(k',z_1,z_2)
\nonumber
\\
&=\theta(x_{12}^0)\partial_{\mathbf{x}_{12}}^{2m}\<\Phi(\mathbf{x}_1,z_1)\Phi(\mathbf{x}_2,z_2)\>_{\text{free}}+(1\leftrightarrow2).
\end{align}
To arrive at the second line, which is exactly \eqref{eq:disp_in_x_space}, we evaluated the trivial $s'_-$ integral and used \eqref{eq:disc_G_to_W} to express the final result in terms of Wightman functions.
We also used that the vacuum is the lowest energy state in the theory,
\begin{align}
\Phi(\mathbf{k},z)|0\>=0, \quad \text{ if } \ k^{0}<0 \ \text{ or } \mathbf{k}^{2}>0.
\end{align}
As a reminder we are using the mostly plus AdS metric.

Finally, we see that the difference between the two amputated diagrams is,
\begin{align}
A^{\text{Feyn}}_\O(\mathbf{x}_{12},z_1,z_2)-A^{\text{disp}}_\O(\mathbf{x}_{12},z_1,z_2)=\left(\partial_{\mathbf{x}_{12}}^{2m}\theta(x_{12}^0)-\theta(x_{12}^0)\partial_{\mathbf{x}_{12}}^{2m}\right)\<[\Phi(x_1,z_1),\Phi(x_2,z_2)]\>.\label{eq:disp_diff}
\end{align}
The right hand side of \eqref{eq:disp_diff} vanishes whenever $(\mathbf{x}_1,z_1)\neq(\mathbf{x}_1,z_2)$.
If the two operators are spacelike separated, then the commutator vanishes.
On the other hand, if the two operators are timelike separated, then the derivatives commute with the $\theta$-functions and \eqref{eq:disp_diff} is again zero.
Therefore, \eqref{eq:disp_diff} is at most proportional to $\delta(\mathbf{x}_{12})\delta(z_1-z_2)$ and its derivatives.
This proves \eqref{eq:APP_single_bulk}, or that the difference between the amputated diagrams is localized at a single bulk point. This in turn implies \eqref{eq:AdS_ambiguities}, which we reproduce below:
\begin{align}
W^{\text{Feyn}}_{\O}(k_a,s_-,t_-)=W^{\text{disp}}_{\O}(k_a,s_-,t_-)+\text{(contact diagrams)}.\label{eq:ambiguitiesAPP}
\end{align}

Note that in this appendix we did not need to use the explicit form of the bulk-to-bulk propagator.
Therefore this argument will also be valid in general asymptotically AdS spacetimes which preserve Poincar\'e invariance of the dual QFT.
For example, we can consider general metrics of the form,
\begin{align}
ds^{2}=e^{2A(z)}(dz^{2}+\eta_{\mu\nu}dx^{\mu}dx^{\nu}),
\end{align}
where $A(z)$ is the warp factor which asymptotes to the AdS warp factor at $z\rightarrow0$ and $z\rightarrow\infty$.
We also did not need to take the radial cutoff $z_c$ to zero.
We did however use that the AdS theory is local, causal and unitary.
For example, we used that the vacuum is stable and that operators commute at spacelike separation.
For this reason, the arguments used in this section only apply directly to exchange diagrams in unitary AdS theories.
That being said, we can extend \eqref{eq:ambiguitiesAPP} to dS exchange diagrams, including those involving heavy fields, by analytically continuing in both $k$ and $\nu$.
It would be interesting to understand if the tree-level dispersion formula \eqref{eq:ambiguities} can also be derived by invoking non-perturbative dS unitarity directly.

\bibliography{Biblio}{}
\bibliographystyle{jhep}

\end{document}